\documentclass[11pt,a4paper]{article}

\usepackage{amsfonts}
\usepackage{multirow}
\usepackage{graphicx}
\usepackage{float}

\title{\textbf{Dissecting zero modes and bound states on \\ BPS vortices in Ginzburg-Landau superconductors.}}

\author{A. Alonso Izquierdo$^{(a)}$, W. Garcia Fuertes$^{(b)}$, and J. Mateos Guilarte$^{(c)}$
\\ {\normalsize {\it $^{(a)}$ Departamento de Matematica
Aplicada}, {\it Universidad de Salamanca, SPAIN}}\\ {\normalsize {\it $^{(b)}$ Departamento de Fisica}, {\it Universidad de Oviedo, SPAIN}} \\ {\normalsize {\it $^{(c)}$ Departamento de Fisica
Fundamental}, {\it Universidad de Salamanca, SPAIN}}}

\date{}

\begin{document}

\maketitle

\begin{abstract}
In this paper the zero modes of fluctuation of cylindrically symmetric self-dual vortices are analyzed and described in full detail. These BPS topological defects arise at the critical point between Type II and Type I superconductors, or, equivalently, when the masses of the Higgs particle and the vector boson in the Abelian Higgs model are equal. In addition, novel bound states of Higss and vector bosons trapped by the self-dual vortices at their core are found and investigated.
\end{abstract}

PACS: 11.15.Kc; 11.27.+d; 11.10.Gh

\section{Introduction}

Vortex filaments carrying a single quantum of magnetic flux were discovered by Abrikosov in the realm of the Ginzburg-Landau theory of Type II superconductors in Reference \cite{Abrikosov}. The same magnetic flux tubes reappeared in the relativistic context of the Abelian Higgs model in the paper of Nielsen and Olesen \cite{Nielsen}, where their stringy nature was emphasized. Analytic formulas are available in this Reference for the $1$-vortex profile near the center of the core and far away from the origin, although the behaviour at infinity was refined in \cite{Perivolaropoulos}. An important step forward in our knowledge of the mathematical properties of these extended structures was achieved by Bogomolny, who identified in the seminal paper \cite{Bogomolny} a system of first-order PDE such that their solutions are the ANO vortices at the transition point between Type II and Type I superconductivity phases, the critical value where the quotient of the scalar and vector particle masses is one. These Bogomolny-Prasad-Sommerfield, \cite{Prasad}, or self-dual{\footnote{Prasad and Sommerfield found magnetic monopoles at the BPS limit of the Georgi-Glashow model where the Higgs potential disappears but still the vacuum orbit is a two-dimensional sphere. \lq\lq Self-dual\rq\rq refers to the fact that the first-order PDE systems, either governing vortices or monopoles, come from two different dimensional reductions of the self-duality Yang-Mills equations.}}, vortices have very interesting features: (1) The magnetic flux is a topological quantity related to the first homotopy group of the circle of degenerate vacua. (2) At the BPS limit, these line defects do not interact with one another. They are thus free to move and zero modes of BPS-vortex fluctuations exist.

The primary aim of this work is to investigate the self-dual vortex zero modes of fluctuation. The BPS vortex PDE equations
admit multivortex solutions. Proof of the existence of this type of solitons was given by Jaffe and Taubes in Reference \cite{Taubes}. The BPS multivortex moduli space with magnetic flux equal to $\frac{2\pi}{e}n$, with $n\in\mathbb{Z}$ integer, is the space of $\vert n\vert$ unordered points in the $\mathbb{R}^2$ plane \cite{Manton}. The freedom in the locations of the centers preludes the existence of $2 \vert n\vert$ linearly independent zero modes of fluctuation, a fact proved by E. Weinberg in \cite{Weinberg} with a shrewd generalization of the index theorem of elliptic operators. This computation, motivated by a physical problem, paved the way to extending the Atiyah-Singer index theorem usually observed in compact spaces, with or without boundaries, to open spaces where problems with the continuous spectrum arise, see e.g. \cite{Callias,Bott}. The self-dual vortex solutions with cylindrical symmetry aroused special interest. In this case the vortex first-order equations reduce to an ordinary differential equation system which is solvable near the origin and very far from the vortex core. Several interpolation methods have been developed, either numerically or through some functional series expansion, to obtain the full multivortex solution, which is never expressible in terms of elementary or special functions, see \cite{de Vega}. In a later development, the self-dual cylindrically symmetric vortex zero mode fluctuations were studied in detail starting with E. Weinberg seminal paper \cite{Weinberg}, see also Chapter 3 in the recent monograph \cite{Weinberg2}. Given the r$\hat{\rm o}$le of the vortex zero modes in the analysis of the low-energy vortex dynamics as geodesic motion in the moduli space of BPS vortex solutions, see e.g. \cite{Manton}-\cite{guilarte}, better ansatzes for the analytical structure of the zero mode fluctuations of BPS vortices were proposed for this purpose in References \cite{Ruback,Burzlaff}. This task was fully achieved in the papers just mentioned for the solutions with a low number of magnetic flux quanta, e.g., $n=2,3$. In the first half of this paper we perform a complete and detailed analysis of the structure of the zero modes of fluctuation around BPS cylindrically symmetric vortices. Relying on the Ruback-Burzlaff ansatz, we describe the vortex zero mode profiles with the same level of precision as the precision attained in the  knowledge of the BPS vortices themselves. After identifying analytically the zero mode radial profile near the core and close to infinity we perform the interpolation between these two regimes by means of a shooting procedure implemented numerically. The angular dependence of the zero mode wave function is fixed analytically by Fourier analysis. The regularity of the wave function near the origin and exponential decay at infinity, all together guaranteeing normalizability, impose the existence of $2 \vert n\vert$ linearly independent vortex zero modes in concordance with the index theorem. The interest of this study is twofold: (1) It extends the work of several authors on this subject to BPS vortices with more than three quanta of magnetic flux. (2) Recently, in \cite{Alonso} and \cite{Alonso1} two of us improved on the one-loop shift calculations of kink masses and domain wall surface tensions by controlling the inaccuracies induced by zero modes in the heat kernel/zeta function regularization procedure. The new method requires precise information about the zero mode wave functions such that the information gathered in this paper is necessary to improve the results obtained in \cite{Marina,Marina1,guilarte1,Marina2,guilarte2} by diminishing the impact of zero modes in heat kernel expansions {\footnote{ Our method applies not only to conventional topological defects but also to instantons, see Reference \cite{alonso2}.}}.

However, vortex zero modes exist and are influential not only in critical vortices between Type I and II superconductors. Jackiw et alli, see e.g. \cite{JackiwR}, discovered that the spectrum of the Dirac operator in a vortex background includes $\vert n\vert$ linearly independent fermionic eigenfunctions of zero eigenvalue, where $n$ is the vortex magnetic charge. From a mathematical point of view the existence of zero modes in the vortex-fermion system obeys an index theorem on a open space, the ${\mathbb R}^2$ plane. Besides these topological roots underlying their existence, the vortex-fermion zero modes add quantum states in the middle of the mass gap of the Dirac spectrum which, in turn, induce the phenomenon of fractionary charge. Thus, the context in which fermionic zero modes in a classical vortex field are considered is completely different: there are no scalar and vector particles and the vortex external field does not fluctuate as in the Abelian Higgs model. Quite recently, this problem gained importance in condensed matter physics, for instance in the mathematics and physics of graphene, see e.g. \cite{JackiwP}, or, in class A chiral superconductors, see \cite{Volovik}.

Soon after the discovery of vortex filaments in Type II superconductors, interest aroused in the investigation of fermionic bound states
trapped at the vortex core by looking at the one-particle spectrum of the Bogolyubov-de Gennes equation near a magnetic flux line background, see \cite{deGennes}. This pioneer paper by de Gennes et al prompted a long search aimed at unveiling the nature of this type of bound states, although without complete success from the analytic point of view. Nevertheless, interesting effects of these bound states on the vortex core have been disclosed in a superfluid phase of the ${}^3He$ isotope, see \cite{Kopnin}. As a secondary goal, we shall study here the bound states arising when scalar and/or vector bosons are trapped at the core of a self-dual vortex in the framework of the Abelian Higgs model, mutatis mutandis in the Ginzburg-Landau phenomenological theory of superconductivity. Contrarily to the bound states mentioned above the particles trapped by the BPS vortices are bosons rather than fermions. In the context of the Abelian Higgs model bound states of mesons by vortices were discovered by Goodman and Hindmarsh in Reference \cite{Hindmarsh} in the mid nineties. Another papers where
the r$\hat{\rm o}$le of these bound states in the framework of topological defects in Cosmology is emphasized are \cite{Arodz1,Arodz2,Kojo}. Taking profit of the supersymmetric quantum mechanical structure linked to BPS topological defects we were able in the short letter \cite{AlGarGuil} to offer a quite detailed description of such bizarre bound states. We shall develop in this work a more complete analysis of the meson bound states on BPS vortices and we shall discuss their properties by comparison with the well known BPS vortex zero modes. Our approach follows the pattern found in the $\lambda \phi^4$ kink. Fluctuations of the domain wall defects in this model are of three types: 1) translational (zero) modes where a meson travels together with the kink center of mass without disturbing the defect profile. 2) kink internal modes of fluctuation where a meson is trapped forming a meson-kink bound state that produces an oscillating in time deformation of the defect profile. The existence of this second type of fluctuations is due to the supersymmetric quantum mechanics of the kink stability problem. 3) Scattering of mesons through the wall.

We shall address types 1) and 2) of fluctuation concerning the BPS vortices in the Abelian Higgs, a much more difficult task. In fact, the search for vortex fluctuations with frequencies greater than 0 but lower than the threshold of the continuous spectrum only differs from the search for zero modes in the fact that the eigenvalue is unsettled a priori. The shooting procedure for obtaining the form factor of the bound state in the intermediate region is thus ineffective and we shall approximate the radial ODE by means of a discretization of the radial coordinate, transforming this ODE into a linear system of difference equations. The
bound state eigenvalues will be identified via diagonalization of the matrix of the linear system, after which the eigenfunctions will be found numerically.

The paper is organized as follows: In Section \S.2 the Abelian Higgs model is revisited with the aim of fixing our notational conventions.
Section \S.3 is devoted to describing in a detailed manner the critical regime between Type I and Type II superconductivity leading to the BPS system of first-order PDE governing the static solutions of finite energy density. Also, the second-order differential operator, which
is usually referred to as the Hessian, determining the small fluctuations around the vortex solutions is discussed in this Section and its factorization as the product of two first-order PD operators is explained. In Sections \S.4 and \S.5 the general structure of the fluctuation spectrum of cylindrically symmetric BPS vortices is developed forming the main contribution of the paper. Section \S.4 offers a comprehensive analysis of the BPS vortex zero modes of magnetic flux $n$ and unveils the general pattern of zero mode fluctuations of a cylindrically symmetric BPS vortex carrying $n$ quanta of magnetic flux. In Section \S.5  a similar picture describing the features of several boson-vortex bound states also with low magnetic charge, is developed. Finally, in Section \S.6 we draw some conclusions and speculate about some future prospects.

\section{ Topological defects carrying quantized magnetic flux in superconducting systems}

We start from the action of the Abelian Higgs model that describes the minimal coupling between a $U(1)$-gauge field and a charged scalar field in a phase where the gauge symmetry is broken spontaneously. In terms of non-dimensional coordinates, couplings and fields, the action functional for this relativistic system in $\mathbb{R}^{1,2}$ Minkowski space-time reads:
\begin{equation}
S[\phi,A]=\int d^3 x \left[ -\frac{1}{4} F_{\mu\nu}F^{\mu \nu} + \frac{1}{2} (D_\mu \phi)^* D^\mu \phi -\frac{\kappa^2}{8} (\phi^* \phi-1)^2 \right]
\label{action1} \qquad .
\end{equation}
The main ingredients are one complex scalar field, $\phi(x)=\phi_1(x)+i\phi_2(x)$, the vector potential $A_\mu(x)=(A_0(x),A_1(x),A_2(x))$, the covariant derivative $D_\mu \phi(x) = (\partial_\mu -i A_\mu(x))\phi(x)$ and the electromagnetic field tensor $F_{\mu\nu}(x)=\partial_\mu A_\nu(x) - \partial_\nu A_\mu(x)$. We choose the metric tensor in Minkowski space in the
form $g_{\mu\nu}={\rm diag}(1,-1,-1),$ with $\mu,\nu=0,1,2$, and use the Einstein repeated index convention. In the temporal gauge $A_0=0$, the energy of static field configurations becomes
\[
E[\phi,A]=\int d^3 x \Big[ \frac{1}{4} F_{ij} F_{ij} + \frac{1}{2} (D_i\phi)^* D_i\phi + \frac{\kappa^2}{8} (\phi^*\phi-1)^2 \Big] \, , \, \, i,j=1,2 \, \,
\]
which, in a non-relativistic context, is the free energy of a superconducting material arising in the Ginzburg-Landau theory of superconductivity, see formula (17) in \cite{Abrikosov} where the order parameter $\phi$ responds to the Cooper pairs density. The search for static configurations requires us to look at the extrema of the functional:
\[
V[\phi,A]= \int\!\!\!\int dx_1dx_2 \Big[ \frac{1}{2} F_{12}^2 + \frac{1}{2} (D_1\phi)^* D_1\phi + \frac{1}{2} (D_2\phi)^* D_2\phi+\frac{\kappa^2}{8} (\phi^*\phi-1)^2 \Big]
\]
The critical points of $V[\phi,A]$ are the static fields satisfying the
second-order PDE system
\begin{eqnarray}
&&(D_1D_1+D_2D_2)\phi(x_1,x_2)=\frac{1}{2}\,\kappa^2\,\phi(x_1,x_2)\, [\phi^*\phi(x_1,x_2)-1] \nonumber \\
&&\partial_2^2 A_1(x_1,x_2)-\partial_2\partial_1 A_2(x_1,x_2)=- \frac{1}{2} \, i \, [\phi^*(x_1,x_2)D_1\phi(x_1,x_2)-\phi(x_1,x_2)(D_1\phi)^*(x_1,x_2) ] \label{pde2}\\
&&\partial_1^2 A_2(x_1,x_2) - \partial_1\partial_2 A_1(x_1,x_2)= \frac{1}{2} i [\phi^*(x_1,x_2) D_2\phi(x_1,x_2)-\phi(x_1,x_2)(D_2\phi)^*(x_1,x_2)] \nonumber .
\end{eqnarray}
Solutions of (\ref{pde2}) that comply with the asymptotic boundary conditions at the circle at infinity, i.e. when $r=\sqrt{x_1^2+x_2^2}\rightarrow\infty$,
\begin{equation}
\phi^* \phi|_{S_\infty^1}=1 \hspace{0.5cm} , \hspace{0.5cm} D_i\phi|_{S_\infty^1}=0 \hspace{0.5cm} \mbox{and} \hspace{0.5cm} F_{12}|_{S_\infty^1}=0 \, \,  \label{asymptotic}
\end{equation}
have finite energy. In fact, choosing
\begin{equation}
\phi\vert_\infty=e^{i n \theta} \hspace{0.8cm} \mbox{and} \hspace{0.8cm} (A_1,A_2 )\vert_\infty=(-i e^{-i n\theta}\partial_1 e^{i n \theta},-i e^{-i n\theta}\partial_2 e^{i n \theta}) \, \, , \label{asymptotic1}
\end{equation}
where $\theta={\rm arctan}\frac{x_2}{x_1}$ and $n$ is an integer, as representatives of (\ref{asymptotic}), one checks that the configuration space
of the static fields is the union of $\mathbb{Z}$ topologically disconnected sectors: ${\cal C}=\left\{ (\phi, A)/ V[\phi,A]<+\infty\right\}=\sqcup_{n\in\mathbb{Z}}\,\,  {\cal C}_n$. The fields in each sector ${\cal C}_n$ are asymptotically constrained by the
formula (\ref{asymptotic1}) where it is evident that $n$ is the winding number of the map $\phi \vert_\infty: \lim_{r\to \infty} S^1_r \, \, \longrightarrow  \, \, S^1$ from the circle at infinity in the $x_1:x_2$-plane to the vacuum orbit determined by the phase of the scalar field at infinity. Thus, all the field configurations in the non-trivial sectors, $n\neq 0$, are endowed with a quantized magnetic flux:
\[
\Phi=\frac{1}{2\pi} \int_{\mathbb{R}^2} d^2 x \, F_{12}=\frac{1}{2\pi}\oint_{S^1_\infty}(A_1dx_1+A_2dx_2)=n \in \mathbb{Z} \, \, .
\]
Rotationally symmetrical solutions of (\ref{pde2}) with finite energy for $\kappa^2\neq 1$ and a quantum of magnetic flux, $n=\pm 1$, are vortices given that the vector field $(A_1,A_2)$ is purely vorticial. Choosing, e.g., $n=1$, the ansatz, see \cite{Nielsen},
\begin{equation}
\phi(x_1,x_2)=f(r)e^{i\theta} \quad , \quad (A_1(x_1,x_2),A_2(x_1,x_2))= \left(-\frac{x_2}{r^2}\,\beta(r), \frac{x_1}{r^2} \, \beta(r)\right)\, , \label{cyan}
\end{equation}
together with the asymptotic conditions, $f(\infty)=1$ and $\beta(\infty)=1$, and the regularity conditions, $f(0)=0$ and $\beta(0)=0$, convert the PDE system (\ref{pde2}) into a second-order ODE system
\begin{eqnarray}
&&\frac{d^2 f}{dr^2}+\frac{1}{r}\frac{df}{dr}-\frac{(1-\beta)^2}{r^2}+\frac{\kappa^2}{2}f(1-f^2)=0 \label{ode2h}\\
&& \frac{d^2\beta}{dr^2}-\frac{1}{r}\frac{d\beta}{dr}+(1-\beta)f^2=0 \label{ode2v} \, ,
\end{eqnarray}
and at the same time guarantees regular behavior at the origin and appropriate fall-off at infinity. There is no available analytical solution to this ODE system. The asymptotic form of these solutions, however, is known, see \cite{Perivolaropoulos}. Linearization of equations (\ref{ode2h}), (\ref{ode2v}) around the scalar and vector fields vacuum values reveals that:
\begin{equation}
\beta(r) \stackrel{r\to \infty}{\simeq} \, 1+c_V \cdot \sqrt{r}e^{-r} \quad , \quad f(r)\stackrel{r\to \infty}{\simeq} \, \left\{\begin{array}{ccc} 1-c_H \cdot \frac{1}{\sqrt{r}} e^{-\kappa r} &\mbox{if}& \kappa<2 \\ 1-c_H \cdot \frac{1}{r} e^{-2 r} &\mbox{if}& \kappa>2 \end{array}\right. \label{asvort}
\end{equation}
where $c_V$ and $c_H$ are integration constants. The full Nielsen-Olesen vortex profiles can only be identified by numerical methods. One might also search for vortex solutions carrying $n$ quanta of magnetic flux. Vector and scalar mesons respectively produce repulsive and attractive forces of Yukawa type between charged objects of the same type. Thus, an effective potential arises prompting vortex solutions of unit magnetic flux to either repel, if $\kappa>1$ (Type II superconductors), or attract each other
when $\kappa<1$ (Type I superconductivity materials). In the first case, the vortices are arranged in a triangular Abrikosov lattice, whereas in Type I superconductors the magnetic flux aggregates on slices piercing the material.

\section{Self-dual/BPS vortices and their fluctuations}

At the transition point $\kappa=1$ between Type I and II superconductors no forces exist between the vortices, which thus, become very special. In order to investigate these critical vortices it is convenient to write $V[\phi,A]$ in the form, see Reference \cite{Bogomolny}:
\begin{eqnarray*}
&& \hspace{-0.8cm} V[\phi,A]=\frac{1}{2}\int_{\mathbb{R}^2} d^2 x \Big[\Big(F_{12}\pm\frac{1}{2}(\phi^*\phi-1)\Big)^2+ \left\vert D_1\phi\pm iD_2\phi\right\vert^2\Big] \mp\\ &&\mp \frac{1}{2}\int_{\mathbb{R}^2} d^2 x \left[F_{12}(\phi^*\phi-1)-i \left\{(D_1\phi)^*D_2\phi-(D_2\phi)^*D_1\phi\right\}\right]+ \frac{\kappa^2-1}{8}\int_{\mathbb{R}^2} d^2 x  \left(1-\phi^*\phi\right)^2
\end{eqnarray*}
where $d^2 x=dx_1dx_2$. Because $-i \left\{(D_1\phi)^*D_2\phi-(D_2\phi)^*D_1\phi\right\}=-F_{12}\phi^*\phi+\partial_1[\phi^* D_2\phi-(D_2\phi)^*\phi]+\partial_2[(D_1\phi)^*\phi-\phi^*D_1\phi]$ we obtain
\[
V[\phi,A]=\frac{1}{2}\int_{\mathbb{R}^2} d^2 x \Big[\Big(F_{12}\pm\frac{1}{2}(\phi^*\phi-1)\Big)^2+\left\vert D_1\phi\pm iD_2\phi\right\vert^2\Big] \pm \frac{1}{2}\int_{\mathbb{R}^2} d^2 x \Big[F_{12} +\frac{\kappa^2-1}{4}\left(1-\phi^*\phi\right)^2 \Big]
\]
up to a total derivative term that integrates to zero over the whole plane if the fields tend to their vacuum values at infinity. The parameter $\kappa$, determined by the $\phi^4$ and electromagnetic couplings as
$\kappa^2=\frac{\lambda}{e^2}$, measures the quotient between the penetration lengths of the scalar and electromagnetic fields in the superconducting medium. Values such that $\kappa^2>1$ characterize Type II superconductors (typically alloys) whereas type I superconductors (metals) correspond to $\kappa^2<1$, as explained above. In the QFT context the parameter $\kappa$ is the quotient between the masses of the Higgs particle, $m_H=\sqrt{\lambda}\, v$, and the vector meson, $m_V=e\, v$, after the Higgs mechanism has taken place giving to the photon a finite mass.

The critical vortices are solutions of the first-order PDE's
\[
D_1\phi \pm i D_2 \phi =0 \hspace{0.5cm},\hspace{0.5cm} F_{12}\pm \frac{1}{2} (\phi^*\phi -1)=0 \, \, ,
\]
which, written in terms of the real and imaginary parts of the complex scalar field $\phi=\phi_1+i\phi_2$, read
\begin{eqnarray}
\partial_1\phi_1 + A_1 \phi_2 \pm [-\partial_2 \phi_2 + A_2 \phi_1] &=&0 \nonumber \\  \partial_1 \phi_2 - A_1 \phi_1\pm [\partial_2 \phi_1 +A_2 \phi_2]&=&0 \label{pde1} \\
F_{12}\pm {\textstyle\frac{1}{2}}(\phi_1^2 + \phi_2^2-1)&=&0 \, \, \nonumber .
\end{eqnarray}
Moreover, the BPS vortices are subjected to the asymptotic conditions (\ref{asymptotic1}). It is clear that at $\kappa^2=1$ the energy of self-dual vortices saturates the Bogomolny topological bound: $V[\phi,A]=\frac{1}{2}\,\vert\int_{\mathbb{R}^2} d^2 x F_{12}\vert=\pi \vert n\vert$
{\footnote{Recovering the physical dimensions, the magnetic flux and the energy per unit length of self-dual vortices would be: $e\Phi=\frac{e}{2\pi}\int_{\mathbb{R}^2}F_{12}=n$, $V[\phi,A]=\pi \vert n\vert v^2$, where $v$ is the vacuum value of the scalar field.}}. Additionally, it may be checked that self-dual vortices also solve the second-order PDE system (\ref{pde2}). Proof of the existence of vorticial solutions of the PDE system (\ref{pde1}) has been developed in Reference \cite{Taubes}. Given a positive integer $n$, there exists a moduli space of self-dual vortices
solving the PDE system (\ref{pde1}) characterized by $2n$ parameters, the centers of the magnetic flux tubes located at the zeroes of the scalar field counted with multiplicity $n_k$, i.e., $n=\sum_{k=1}^{n_0} \Phi_k n_k$, where $\Phi_k$ is the quantized magnetic flux of a vortex (the multiplicity) and $n_0$ is the total number of flux lines, see also Reference \cite{Manton}. Behind this particular structure lies the fact that there are no forces between self-dual vortices ($\kappa=1$) of one or several quanta of magnetic flux, which thus move freely throughout the $x_1$-$x_2$-plane.

\subsection{The first-order fluctuation operator: hidden Supersymmetric Quantum Mechanics}

Knowing that the energy of self-dual vortices is a topological quantity, there is no doubt about the stability of these topological solitons. The main theme of this paper, however, is the analysis of field fluctuations around self-dual vortices. We shall concentrate on two special modes of fluctuations: 1) the $2 n$ vortex zero modes, those belonging to the kernel of the second-order fluctuation operator (the Hessian in variational calculus terminology), which arise because of the freedom of motion of the centers. 2) vortex internal modes of fluctuation corresponding to bound state normalizable eigenfunctions of the Hessian in the discrete spectrum and 3) scattering eigenfunctions in the continuous spectrum.

Let us denote the scalar field $\phi_V$ and the vector potential $A_V$ corresponding to a self-dual vortex solution of vorticity $n$ as:
\[
\phi_V=\psi(\vec{x};n)=\psi_1(\vec{x};n) + i \, \psi_2(\vec{x};n) \hspace{0.5cm},\hspace{0.5cm} A_V=(V_1(\vec{x};n),V_2(\vec{x};n)) \hspace{0.5cm}\mbox{with}\hspace{0.5cm} \vec{x}=(x_1,x_2)\quad .
\]
The self-dual vortex fluctuations $(a_1(\vec{x}),a_2(\vec{x}))$ and $\varphi(\vec{x})=\varphi_1(\vec{x})+i\varphi_2(\vec{x})$ built around the BPS vortex fields
\begin{eqnarray}
&& \hspace{-3cm}(A_1(\vec{x};n),A_2(\vec{x};n))= (V_1(\vec{x};n),V_2(\vec{x};n) ) +\epsilon \, (a_1(\vec{x}),a_2(\vec{x})) \nonumber\\ && \hspace{-3cm}\phi_1(\vec{x};n)=\psi_1(\vec{x};n) + \epsilon\, \varphi_1(\vec{x}) \quad , \quad \phi_2(\vec{x};n)=\psi_2(\vec{x};n) + \epsilon\, \varphi_2(\vec{x}) \label{perturbed}
\end{eqnarray}
are zero modes of fluctuation if the perturbed fields  (\ref{perturbed}) are still solutions of the first-order equations (\ref{pde1}). To discard pure gauge fluctuations, we select the \lq\lq background\rq\rq{} gauge
\begin{equation}
B(a_k,\varphi,\phi_V)=\partial_k a_k( \vec{x})-(\,\psi_1( \vec{x})\, \varphi_2( \vec{x})-\psi_2( \vec{x})\,\varphi_1( \vec{x})\,)=0
\label{backgroundgauge}
\end{equation}
as the gauge fixing condition on the fluctuation modes. The system of four PDE equations (\ref{pde1})-(\ref{backgroundgauge}) is satisfied if and only if the four-vector field
\[
\xi(\vec{x})=\left( \begin{array}{c c c c}a_1(\vec{x}) & a_2(\vec{x}) & \varphi_1(\vec{x}) & \varphi_2(\vec{x}) \end{array} \right)^t
\]
which assembles all the BPS vortex field fluctuations is annihilated by the first-order PDE operator:
\begin{equation}
{\cal D}= \left( \begin{array}{cccc}
-\partial_2 & \partial_1 & \psi_1 & \psi_2 \\
-\partial_1 & -\partial_2 & -\psi_2 & \psi_1 \\
\psi_1 & -\psi_2 & -\partial_2 + V_1 & -\partial_1 -V_2 \\
\psi_2 & \psi_1 & \partial_1+V_2 & -\partial_2 + V_1
\end{array} \right)  \label{zeromoded} \, \, .
\end{equation}
Note that this ${\cal D}$ operator is obtained by deforming the PDE system (\ref{pde1}) together with the background gauge.
Thus, the zero mode $\xi_0(\vec{x})$ BPS vortex fluctuation fields belong to the kernel of the first-order PDE operator, ${\cal D}\xi_0(\vec{x})=0$, or, in components, are $L^2$ solutions of the PDE system:
\begin{eqnarray}
-\partial_2 a_1 + \partial_1 a_2 +\psi_1 \varphi_1 + \psi_2 \varphi_2 &=& 0 \nonumber \\
-\partial_1 a_1 - \partial_2 a_2 -\psi_2 \varphi_1 + \psi_1 \varphi_2 &=& 0 \label{zeromode2}\\
\psi_1 a_1 -\psi_2 a_2 + (-\partial_2 +V_1)\varphi_1+(-\partial_1-V_2)\varphi_2&=&0 \nonumber \\
\psi_2 a_1 +\psi_1 a_2 + (\partial_1 +V_2)\varphi_1+(-\partial_2+V_1)\varphi_2&=&0 \nonumber \, \, .
\end{eqnarray}
Important information about the particle spectrum in the Abelian Higgs model not only comes from the vortex zero mode fluctuations but also
from vortex fluctuations demanding positive energy because these positive perturbations determine the dynamics of mesons in the different topological sectors. Thus, we shall investigate the spectral condition ${\cal H}^+ \xi_\lambda(\vec{x}) =\omega_\lambda^2 \, \xi_\lambda(\vec{x})$, where $\lambda$ is a label in either the discrete or the continuous spectrum useful to enumerate the eigenfunctions and eigenvalues, and ${\cal H}^+$ is the second-order vortex small fluctuation operator
\[
{\cal H}^+= \left( \begin{array}{cccc}
-\Delta + |\psi|^2 & 0 & -2D_1 \psi_2 & 2 D_1 \psi_1 \\
0 & -\Delta +|\psi|^2 & -2 D_2 \psi_2 & 2 D_2 \psi_1 \\
-2 D_1 \psi_2 & -2 D_2\psi_2 & -\Delta +\frac{1}{2} (3|\psi|^2-1)+V_kV_k & -2 V_k \partial_k -\partial_k V_k \\
2D_1\psi_1 & 2 D_2 \psi_1 & 2V_k \partial_k + \partial_k V_k & -\Delta +\frac{1}{2} (3|\psi|^2-1) + V_kV_k
 \end{array} \right) \quad
\]
coming from linearizing the field equations (in the background gauge) around the BPS vortices, see \cite{guilarte1}. The fluctuation vectors $\xi(\vec{x})$ belong in general to a rigged Hilbert space, such that there exist square integrable eigenfunctions $\xi_j(\vec{x})\in L^2(\mathbb{R}^2)\oplus \mathbb{R}^4$ belonging to the discrete spectrum, for which the norm $\|\xi(\vec{x})\|$ is bounded:
\begin{equation}
\|\xi(\vec{x})\|^2  = \int_{\mathbb{R}^2} d^2x \Big[ (a_1(\vec{x}))^2 + (a_2(\vec{x}))^2 + (\varphi_1(\vec{x}))^2 + (\varphi_2(\vec{x}))^2 \Big] < +\infty \, ,
\label{normalization}
\end{equation}
together with continuous spectrum eigenfunctions $\xi_\nu(\vec{x})$ with $\nu$ ranging in a dense set.
One observes that the second-order differential operator ${\cal H}^+$ is one of the two SUSY partner operators obtained from ${\cal D}$ as
\begin{equation}
{\cal H}^+={\cal D}^\dagger \, {\cal D} \qquad , \qquad {\cal H}^-={\cal D} \, {\cal D}^\dagger \label{susy01}
\end{equation}
which are isospectral in the positive part of the spectrum{\footnote{The continuous part of the spectrum might cause some difficulties, which are explained in \cite{Weinberg}}. Explicitly, we find
\[
{\cal H}^- =  \left( \begin{array}{cccc}
-\Delta + |\psi|^2 & 0 & 0 & 0 \\
0 & -\Delta +|\psi|^2 & 0 & 0 \\
0 & 0 & -\Delta +\frac{1}{2} (|\psi|^2+1)+V_kV_k & -2 V_k \partial_k -\partial_k V_k \\
0 & 0 & 2V_k \partial_k + \partial_k V_k & -\Delta +\frac{1}{2} (|\psi|^2+1) + V_kV_k
 \end{array} \right) \, \, .
\]
The Hamiltonians ${\cal H}^\pm$ are superpartners in a supersymmetric quantum mechanical system built from the \lq\lq supercharges\rq\rq {\footnote{Although written in the text as $2\times 2$-matrices, both the supercharges and the SUSY Hamiltonian are $8\times 8$-matrices of partial differential operators.}}
\[
Q=\left(\begin{array}{cc} 0 & 0 \\ {\cal D} & 0 \end{array}\right) \, \, \, \, \, \quad , \, \, \, \, \, \quad  Q^\dagger=\left(\begin{array}{cc} 0 & {\cal D}^\dagger \\ 0 & 0 \end{array}\right) \quad ,
\]
which is governed by the SUSY Hamiltonian:
\[
{\cal H}=Q Q^\dagger+Q^\dagger Q=\left(\begin{array}{cc} {\cal H}^+ & 0 \\ 0 & {\cal H}^- \end{array}\right) \, \, .
\]

The stability of the BPS $n$-vortex solutions implies that the ${\cal H}^+$-spectrum consists of non-negative eigenvalues.
Indeed in \cite{Weinberg} within the framework of index theory in open spaces, see \cite{Callias}-\cite{Bott}, E. Weinberg proved that there are $2 n$ linearly independent normalizable BPS vortex zero modes in the topological sector of magnetic flux $n$, i.e., the dimension of the
algebraic kernel of ${\cal D}$, henceforth of ${\cal H}^+$ is $n$ .
By inspection one sees that ${\cal H}^-$ lacks zero modes (all the potential wells are non-negative) and the index theorem dictates:
\[
{\rm ind}\,{\cal D} ={\rm dim} \, {\rm Ker} \, {\cal D}=\lim_{M\to \infty} {\rm Tr}_{L^2}\Big\{\frac{M^2}{{\cal D}^\dagger{\cal D}+M^2}-\frac{M^2}{{\cal D}{\cal D}^\dagger+M^2}\Big\}=2 n \, \, ,
\]
which means that ${\cal H}^+$ has $2 n$ zero modes, see Appendix B in \cite{Weinberg2}. Analysis of the zero mode eigenfunctions begun in \cite{Weinberg} and was further developed in References \cite{Ruback} and \cite{Burzlaff}. In the last two references the motivation to describe in detail the vortex zero modes came from the study of vortex scattering at low energies within the approach of geodesic dynamics in their moduli space, see e.g. \cite{guilarte}.

The first goal in this paper is to seek a complete description of the vortex zero modes not fully developed in the previous references because interest was focused in sectors of very low magnetic charge. The second task that we envisage is the search for the eigenfunctions in the strictly positive spectrum of ${\cal H}^+$, in particular the bound states, i.e., the internal modes of fluctuation where the self-dual vortex captures scalar and/or vector mesons.

\section{Zero mode fluctuations of BPS cylindrically symmetric vortices}

The Nielsen-Olesen ansatz (\ref{cyan}) generalized to the topological sector ${\cal C}_n$, in cylindrical coordinates also in field space, reads:
\begin{equation}
\phi(\vec{x})=f_n(r) \, e^{in\theta} \hspace{0.5cm};\hspace{0.5cm} r A_\theta(r,\theta)= n \, \beta_n(r) \label{cyan2} \, \, .
\end{equation}
In this ansatz we assume the radial gauge $A_r=0$, such that the vector field is purely vorticial. Plugging (\ref{cyan2}) into the first-order PDE system (\ref{pde1}), the following ODE system emerges:
\begin{equation}
\frac{df_n}{d r}(r)=\frac{n}{r} f_n(r) [1-\beta_n(r)]\hspace{0.5cm},\hspace{0.5cm} \frac{d\beta_n}{d r}(r)=\frac{r}{2n}[1-f_n^2(r)] \, . \label{ode1}
\end{equation}
The solutions for the radial profiles $f_n(r)$ and $\beta_n(r)$ are the self-dual vortex solutions {\footnote{Without loss of generality, we restrict $n$ and the signs in the first-order system (\ref{pde1}) to be positive.}}. The asymptotic conditions (\ref{asymptotic}) demand that $f_n(r)\rightarrow 1$ and $\beta_n(r)\rightarrow 1$ as $r\rightarrow \infty$. In fact, it is immediate to check that the asymptotic behaviour fits formula (\ref{asvort}) when $\kappa=1$, i.e., the critical value where the mass of the Higgs boson is equal to (henceforth less than twice) the mass of the vector boson. The requirement of regularity at $r=0$ also fixes the behaviour of the solutions near the origin to be $f_n(r)=d_n r^n$ and $\beta_n(r)=e_2 r^2$, where $d_n$ and $e_2$ are integration constants. Choice of these constants must be tailored to fit the asymptotic behaviour. A shooting procedure implemented numerically allows us to solve the system (\ref{ode1}) by interpolating the field profiles between their shapes in the neighborhoods of the origin and of infinity. In this way we construct the cylindrically symmetric self-dual $n$-vortices.

\noindent Investigation of zero mode fluctuations around cylindrically symmetric self-dual vortices begins with rewriting the PDE system (\ref{zeromode2}) in polar coordinates:
\begin{eqnarray}
\frac{\partial a_\theta}{\partial r} - \frac{1}{r} \frac{\partial a_r}{\partial \theta} + \frac{1}{r} a_\theta + f_n(r) \cos(n\theta) \, \varphi_1 + f_n(r) \sin(n\theta)\, \varphi_2&=&0 \label{zeromode31} \\
-\frac{1}{r} \,\frac{\partial a_\theta}{\partial \theta} - \frac{\partial a_r}{\partial r} - \frac{1}{r} \, a_r - f_n(r) \sin (n\theta)\, \varphi_1 + f_n(r) \cos(n\theta) \, \varphi_2&=&0 \label{zeromode32} \\
-\frac{1}{r} \frac{\partial \varphi_1}{\partial \theta} - \frac{\partial \varphi_2}{\partial r} - \frac{n \, \beta_n(r)}{r} \, \varphi_2 + f_n(r) \cos(n\theta) \, a_r - f_n(r) \sin (n\theta) \, a_\theta &=&0 \label{zeromode33} \\
-\frac{\partial \varphi_1}{\partial r} + \frac{1}{r} \frac{\partial \varphi_2}{\partial \theta} - \frac{n\,\beta_n(r)}{r} \, \varphi_1 - f_n(r) \sin(n\theta) \, a_r -f_n(r) \cos(n\theta) \, a_\theta&=&0 \label{zeromode34} \, \, ,
\end{eqnarray}
where we recall that $a_1=a_r \cos \theta - a_\theta \sin \theta$ and $a_2=a_r \sin \theta + a_\theta \cos \theta$.

\subsection{Analytical investigation of the algebraic kernel of the first-order operator ${\cal D}$}

\noindent In this Section we shall prove a proposition characterizing the analytical behavior of the eigenfunctions which belong to the kernel of the operator ${\cal D}$ acting on self-dual rotationally symmetric vortices. First, we state a helpful Lemma which unveils the existence of a useful symmetry in BPS vortex fluctuation space.

\vspace{0.2cm}

\noindent \textsc{Lemma:} Let us assume that $\xi_0(\vec{x})=(a_r,a_\theta,\varphi_1,\varphi_2)^t$ is a zero mode of the second-order  fluctuation operator ${\cal H}^+$. Then, $\xi_0^\perp(\vec{x})=(a_\theta, -a_r,\varphi_2,-\varphi_1)^t$ is a second zero mode of ${\cal H}^+$, which is orthogonal and linearly independent of  $\xi_0(\vec{x})$.

\vspace{0.1cm}

\noindent \textsc{Proof:} The $\frac{\pi}{2}$-rotation in the internal space of scalar and vector field fluctuation planes, $\varphi_1 \rightarrow \varphi_2$, $\varphi_2 \rightarrow -\varphi_1$, $a_r \rightarrow a_\theta$ and $a_\theta \longrightarrow - a_r$, is a symmetry of the ODE system (\ref{zeromode31})--(\ref{zeromode34}) because it replaces (\ref{zeromode31})
by (\ref{zeromode32}) and (\ref{zeromode33}) with (\ref{zeromode34}). This discrete rotation also transforms $\xi_0(\vec{x})$ into $\xi_0^\perp(\vec{x})$. Thus, if $\xi_0(\vec{x})$ belongs to the kernel of ${\cal D}$ evaluated on self-dual cylindrically symmetric vortices, $\xi_0^\perp(\vec{x})$ is also a zero mode around the same vortex solution. The orthogonality between $\xi_0(\vec{x})$ and $\xi_0^\perp(\vec{x})$ follows directly from the $L^2(\mathbb{R}^2)\oplus\mathbb{R}^4$-inner product. $\Box$

\vspace{0.2cm}

\noindent Regarding vortex zero modes the main result is as follows:

\vspace{0.2cm}

\noindent \textsc{Proposition:} There exist $2 n$ orthogonal zero mode $L^2(\mathbb{R}^2)\oplus\mathbb{R}^4 $-fluctuations of the self-dual cylindrically symmetric $n$-vortex solution of the form
\begin{eqnarray}
\xi_0(\vec{x},n,k)&=& r^{n-k-1} \left( \begin{array}{c} h_{nk}(r) \, \sin[(n-k-1)\theta] \\ h_{nk}(r) \, \cos[(n-k-1)\theta] \\  -\frac{h_{nk}'(r)}{f_n(r)} \, \cos(k\theta) \\ - \frac{h_{nk}'(r)}{f_n(r)} \, \sin(k\theta) \end{array} \right) \quad , \label{zeromode4}
\\
\xi_0^\perp(\vec{x},n,k)&=& r^{n-k-1} \left( \begin{array}{c} h_{nk}(r) \, \cos[(n-k-1)\theta] \\ -h_{nk}(r) \, \sin[(n-k-1)\theta] \\  - \frac{h_{nk}'(r)}{f_n(r)} \, \sin(k\theta) \\  \frac{h_{nk}'(r)}{f_n(r)} \, \cos(k\theta)  \end{array} \right) \quad , \label{zeromode41}
\end{eqnarray}
where $k=0,1,2,\dots,n-1$, and the zero mode radial form factor $h_n(r)$ satisfies the second-order ODE
\begin{equation}
-r \, h_{nk}''(r)+[1+2k-2n\,\beta_n(r)]\,h_{nk}'(r) + r \,f_n^2(r)\, h_{nk}(r)=0 \label{ode5}
\end{equation}
with the contour conditions $h_{nk}(0)\neq 0$ and $\lim_{r\rightarrow \infty} h_{nk}(r) =0$.

\vspace{0.1cm}

\noindent \textsc{Proof:} Because the discrete symmetry explained in the previous Lemma we are only interested in the construction of the $n$ zero modes $\xi_0(\vec{x},n,k)$, their orthogonal partners $\xi_0^\perp(\vec{x},n,k)$ follow immediately.

The structure of the PDE system (\ref{zeromode31})-(\ref{zeromode34}) suggests the ansatz
\begin{eqnarray*}
a_r(r,\theta)=g_{nk}(r) \sin[(n-k)\theta] \qquad &;& \qquad  a_\theta(r,\theta)=g_{nk}(r) \cos[(n-k)\theta] \\ \varphi_1(r,\theta)=t_{nk}(r) \cos(k\theta) \qquad &;& \qquad \varphi_2(r,\theta)=t_{nk}(r) \sin(k\theta) \quad ,
\end{eqnarray*}
where the radial and angular dependencies of the components of $\xi_0(\vec{x},n,k)$ are separated, in the search for the BPS vortex zero modes. Plugging this ansatz into the set (\ref{zeromode31})-(\ref{zeromode34}) of four ODE's, only two independent but coupled first-order ODE's remain:
\begin{eqnarray}
&& r \,\frac{d g_{nk}}{dr}(r)+r \,f_n(r) \, t_{nk}(r)+(1+k-n)\,g_{nk}(r)=0 \label{ode21} \\ &&
r \, \frac{d t_{nk}}{dr}(r) + [n \, \beta_n(r)-k]\, t_{nk}(r) + r \, f_n(r) \, g_{nk}(r) =0 \label{ode22} \, \, ,
\end{eqnarray}
which govern the behaviour of the $g_{nk}(r)$ and $t_{nk}(r)$ functions. Continuity in the angular part of the solution requires $k$ to be an integer, $k\in \mathbb{Z}$, and the $L^2$ integrability of the fluctuations demands that
\begin{equation}
\|\xi_0(\vec{x},n,k)\|^2 = 2\pi \int r \, dr \Big[ g_{nk}^2(r) + t_{nk}^2(r) \Big] \, < \, +\infty
\label{normalization3}
\end{equation}
Solving for the function $t_{nk}(r)$ in (\ref{ode21}), we obtain
\begin{equation}
t_{nk}(r)=-\frac{g_{nk}'(r)}{f_n(r)} - \frac{(1+k-n)g_{nk}(r)}{r\,f_n(r)} \, \, , \label{tnr}
\end{equation}
and plugging this expression into (\ref{ode22}) we end with a single second-order ODE for $g_{nk}(r)$
\begin{equation}
-r^2 g_{nk}''(r)- r\,[1-2n +2n\,\beta_n(r)\,] \,g_{nk}'(r) + \left[(1+k-n)(1+k+n-2n\, \beta_n(r))+r^2 f_n^2(r)\right] \, g_{nk}(r)=0 \label{ode3}
\end{equation}
in terms of the self-dual vortex profiles $f_n(r)$ and $\beta_n(r)$. Now we shall investigate the behaviour of this function:

\begin{itemize}
\item \textit{Regularity of the function $g_{nk}(r)$ at the origin:} The origin $r=0$ is a regular singular point of the second-order differential equation (\ref{ode3}). In this situation there exists a single analytic solution at $r=0$ although the second linearly independent solution of (\ref{ode3}) has a singularity. The analytic solution admits a series expansion around $r=0$ of the form:
\begin{equation}
g_{nk}(r)=r^s \sum_{j=0}^\infty c_{j}^{(n,k)}\, r^j = r^s \, h_{nk}(r) \hspace{0.5cm},\hspace{0.5cm} h_{nk}(r)= \sum_{j=0}^\infty c_j^{(n,k)}\, r^j \, . \label{frob}
\end{equation}
$s$ is chosen as the minimum value that selects $c_{0}^{(n,k)}\neq 0$, i.e., $h_{nk}(r)$ is regular, and does not vanish, at $r=0$. Plugging (\ref{frob}) into (\ref{ode3}), and taking into account that $f_n(r) \sim r^n \sum_{\ell=0}^\infty d_{n+2\ell}\, r^{2\ell}$ and $\beta_n(r)\sim e_2 r^2 + r^{2n+2}\sum_{\ell=0}^\infty e_{2n+2+2\ell} \, r^{2\ell}$ near the origin, we obtain the identity
\begin{eqnarray}
&& \sum_{j=0}^\infty \Big[-(-1+j-k-n+s)(1+j+k-n+s) \, c_{j}^{(n,k)} \, r^j \Big] + \nonumber\\ &&+ \sum_{j=2}^\infty \Big[ -2 \, n \, e_2 \, (-1+j+k-n+s)\, c_{j-2}^{(n,k)} \, r^{j} \Big] + {\cal O}(r^{2n+1})=0 \label{recurrencia1}
\end{eqnarray}
from which we extract recurrence relations that determine the series expansion coefficients $c_{j}^{(n,k)}$ up to order $2n+1$. Annihilation of the term independent of $r$ in (\ref{recurrencia1}) unveils the indicial equation
\begin{equation}
(1+k+n-s)(1+k-n+s)\, c_{0}^{(n,k)}=0 \label{ind} \, \, .
\end{equation}
Because $c_{0}^{(n,k)}\neq 0$, the two characteristic exponents, the two values of $s$ compatible with (\ref{ind}), are:
\[
\mbox{A:}\hspace{0.4cm} s=n-k-1 \hspace{1cm} \mbox{or} \hspace{1cm} \mbox{B:}\hspace{0.4cm} s=n+k+1 \quad .
\]
Both possibilities are equivalent: simply redefine $k$, $k\leftrightarrow -(k+2)$. Thus, we shall stick to choice $A$. In terms of the function $h_{nk}(r)$, let us recall that $g_{nk}(r)=r^{n-k-1} \, h_{nk}(r)$, the norm (\ref{normalization3}) of the vortex zero modes is
\begin{equation}
\|\xi_0(\vec{x},n,k)\|^2 =2\pi\int_0^\infty \, rdr \, \rho_{nk}(r)=2\pi \int_0^\infty dr \, r^{2(n-k)-1} \Big[ h_{nk}^2(r) + \frac{(h_{nk}'(r))^2}{f_n^2(r)} \Big] .
\label{normalization2}
\end{equation}
Near the origin, the behaviour of the first summand of the integrand of (\ref{normalization2}) is
\[
r^{2(n-k)-1} \,h_{nk}^2(r) \approx (c_{0}^{(n,k)})^2 \, r^{2(n-k)-1} + {\cal O}(r^{2(n-k)+1}) .
\]
In order to avoid singularities in the integrand coming from a pole at the origin, we demand that
\[
2(n-k)-1\geq 0 \hspace{0.5cm} \Rightarrow \hspace{0.5cm}  k\leq n-1 \quad .
\]
The term proportional to $r$ in (\ref{recurrencia1}) is null if $(1+2k)c_{1}^{(n,k)}=0$ for the characteristic exponent A. Thus, the second coefficient in the expansion vanishes: $c_{1}^{(n,k)}=0$. Moreover, the recurrence relations extracted from (\ref{recurrencia1}) for the odd  terms proportional to $r^{2 i+1}$, with $i=1,2,\dots, [n-\frac{1}{2}]$,
\[
(2i+1)(2k-2i+1)\,c_{2i+1}^{(n,k)}=2 e_2 (2i-1) \, n \, c_{2i-1}^{(n,k)}
\]
show that all the odd coefficients $c_{2 i +1}^{(n,k)}$ also vanish, at least up to $c_{2n+1}^{(n,k)}$, given that $c_{1}^{(n,k)}=0$ and the constants multiplying $c_{2i+1}^{(n,k)}$ and $c_{2i-1}^{(n,k)}$ in the equation above are non-null. The two-term recurrence relations between the even coefficients is read from the annihilation of the terms proportional to $r^{2 i}$, $i=1,2,\dots,n$, in (\ref{recurrencia1}):
\begin{equation}
i\,(k-i+1)\,c_{2i}^{(n,k)}=e_2\,(i-1)\,n\, c_{2i-2}^{(n,k)} \quad . \label{erec}
\end{equation}
The relations (\ref{erec}) imply $c_{2}^{(n,k)}=c_{4}^{(n,k)}=c_{6}^{(n,k)}=\cdots =0$ up to the $c_{2k}^{(n,k)}$ coefficient. When $i=k+1$ the left member vanishes despite $c_{2k+2}^{(n,k)}$ not being null. Thus, in a neighborhood of the origin the function $h_{nk}(r)$ adopts the form:
\begin{equation}
h_{nk}(r)=c_{0}^{(n,k)} + c_{2k+2}^{(n,k)} \,r^{2k+2} + \dots \label{hcerca0}
\end{equation}
with $c_{0}^{(n,k)}$ and $c_{2k+2}^{(n,k)}$ arbitrary constants. Therefore, near the origin the second summand of the integrand in (\ref{normalization2}) behaves as:
\[
r^{2(n-k)-1} \, \frac{(h_{nk}'(r))^2}{f_n^2(r)} \approx \, (c_{2k+2}^{(n,k)})^2 \, (2k+2)^2\, r^{2k+1} + {\cal O}(r^{2k+3})
\]
Singularities in the integrand coming from a pole at the origin are prevented if the inequality
\[
2k+1 \geq 0 \hspace{0.5cm} \Rightarrow \hspace{0.5cm} k\geq 0
\]
holds. Together with the inequality $k\leq n-1$, the univaluedness of the BPS vortex zero modes $\xi_0(r,\theta, n, k)=\xi_0(r,\theta+ 2\pi, n, k)$, equivalent to $k\in \mathbb{Z}$ being an integer, restricts the possible values of $k$ to being $0\leq k \leq n-1$, as stated in the proposition.

\item \textit{Asymptotic behaviour of the function $g_{nk}(r)$:} The following step is to investigate the asymptotic behaviour of the function $g_{nk}(r)$ far from the origin. Replacing the asymptotic vortex profiles when $r\to +\infty$, $f_n(r)\rightarrow 1$ and $\beta_n(r)\rightarrow 1$ into the differential equation (\ref{ode3}) the modified Bessel equation
\[
-r^2 g_{nk}''(r)-rg_{nk}'(r)+[(1+k-n)^2+r^2]g_{nk}(r)=0
\]
appears. The solution is a linear combination of the modified Bessel functions with very well known asymptotic properties:
\begin{equation}
g_{nk}(r) \stackrel{r\rightarrow +\infty}{\simeq}  C_1 I_{1+k-n}(r)+C_2 K_{1+k-n}(r) \stackrel{r\rightarrow +\infty}{\simeq} C_1 \frac{e^r}{\sqrt{r}} + C_2 \frac{e^{-r}}{\sqrt{r}}  \label{asymptotic2}\,\, .
\end{equation}
\end{itemize}

\noindent By adequately tuning the integration constants $c_0^{(n,k)}$ and $c_{2k+2}^{(n,k)}$ we can get the exponentially decaying particular solution. We thus keep the exponentially decaying tail (\ref{asymptotic2}) in the vortex zero mode eigenfunctions that does not impose new restrictions. We conclude the finiteness of the norm (\ref{normalization2}) of $\xi_0(\vec{x};n,k)$ and therefore $\xi_0(\vec{x},n,k)\in L^2(\mathbb{R}^2)\oplus\mathbb{R}^4$. Together with their partners $\xi_0^\perp(\vec{x};n,k)$, perpendicular to them in field space, they form a system of $2n$ orthogonal zero modes of fluctuation around any rotationally symmetric self-dual vortex of magnetic flux $n$ that are normalizable. The orthogonality between the zero modes $\xi_0(\vec{x};n,k)$ (\ref{zeromode4}) arises directly from the angular dependence of these eigenfunctions. Now let us insert $g_{nk}(r)=r^{n-k-1} h_{nk}(r)$ into equations (\ref{ode3}); we end with the differential equation (\ref{ode5}) and the general form (\ref{zeromode4}) of the vortex zero modes according to the above Proposition. Notice that $t_{nk}(r)=-r^{n-k-1}\frac{h_{nk}'(r)}{f_n(r)}$. $\Box$

\subsection{Weakly deformed cylindrically symmetric BPS vortices: moving around magnetic flux quanta}

We have just found new solutions of the self-duality equations in the linear approximation obtained by means of perturbations of the BPS cylindrically symmetric vortices induced by adding to them the $\xi_0(\vec{x};n,k)$ zero modes, mutatis mutandis the $\xi_0^\perp(\vec{x};n,k)$ zero modes. Therefore, the following infinitesimal deformations of the self-dual rotationally symmetric $n$-vortex are still self-dual solutions of the vortex equations:
\begin{eqnarray}
\widetilde{\psi}_1(\vec{x};n,k)&=& \psi_1(\vec{x},n) + \epsilon \, \varphi_1(\vec{x};n,k) =\psi_1(\vec{x},n) - \epsilon \, r^{n-k-1} {\textstyle\frac{h_{nk}'(r)}{f_n(r)}} \cos(k\theta) \nonumber \\
\widetilde{\psi}_2(\vec{x};n,k) &=& \psi_2(\vec{x},n) + \epsilon \, \varphi_2(\vec{x};n,k) =\psi_2(\vec{x},n) - \epsilon \, r^{n-k-1} {\textstyle\frac{h_{nk}'(r)}{f_n(r)}} \sin(k\theta) \label{psdhf} \\ \widetilde{V}_1(\vec{x};n,k)&=& V_1(\vec{x},n)+\epsilon \, a_1(\vec{x};n,k)=V_1(\vec{x},n)+\epsilon \, r^{n-k-1} h_{nk}(r)\sin[(n-k-1)\theta] \nonumber\\ \widetilde{V}_2(\vec{x};n,k)&=& V_2(\vec{x},n)+\epsilon \, a_2(\vec{x};n,k)=V_2(\vec{x},n)+\epsilon \, r^{n-k-1} h_{nk}(r)\cos[(n-k-1)\theta] \nonumber \, \, ,
\end{eqnarray}
or, in complex variables in field space, $\psi=\psi_1+i\psi_2$ and $V=V_1+iV_2$,
\begin{eqnarray*}
\widetilde{\psi}(\vec{x};n,k)&=& f_n(r)e^{in\theta} - \epsilon \, r^{n-k-1}\frac{h_{nk}'(r)}{f_n(r)}\,  e^{i k\theta}\\
\widetilde{V}(\vec{x};n,k)&=& i\frac{n}{r^2} \, \beta_n(r) \, e^{i \theta}+i \, \epsilon \, r^{n-k-1}h_{nk}(r) \, e^{i(k+1-n)\theta} \quad .
\end{eqnarray*}
The main properties of the perturbed self-dual solutions are unveiled by studying the behaviour of the Higgs field near the origin:
\[
\widetilde{\psi}(\vec{x},n,k)=d_n r^n e^{in\theta} - \epsilon \, \frac{(2k+2)c_{2k+2}^{(n,k)}}{d_n} r^{k} e^{i k\theta} =  r^k e^{ik\theta} \Big[ d_n r^{n-k} e^{i(n-k)\theta} - \epsilon \, \frac{(2k+2)c_{2k+2}^{(n,k)}}{d_n} \Big]
\]
which tells us that the origin is a zero of $\widetilde{\psi}(\vec{x},n,k)$ with multiplicity $k$ and that other $n-k$ zeroes appear characterized by the conditions
\[
r^{n-k}= \epsilon \frac{(2k+2)\vert c_{2k+2}^{(n,k)}\vert}{d_n^2} \hspace{1.5cm} \mbox{\underline{and}} \hspace{1.5cm} e^{i(n-k)\theta}={\rm sign} ( c_{2k+2}^{(n,k)})=-1 \, \, ,
\]
where we have used the fact that $h_{nk}(r)$ is a decreasing function near the origin. Therefore, $n-k$ zeroes located at the origin in the rotationally symmetric self-dual vortices migrate under this perturbation to
the positions
\[
r_0(n,k,j)e^{i\theta(n,k,j)}=\Big[\epsilon \frac{(2k+2)\vert c_{2k+2}^{(n,k)}\vert}{d_n^2} \Big]^\frac{1}{n-k} \, e^{i\frac{(2j+1)\pi}{n-k} } \hspace{0.5cm}, \hspace{0.5cm} j=0,1,\dots,n-k-1
\]
infinitesimally apart from the origin. In sum, the perturbation induced by the zero mode $\xi_0(\vec{x};n,k)$ on a self-dual $n$-vortex leaves $k$ quanta of magnetic flux at the origin but moves $n-k$ Higgs field zeroes away to be placed on the vertices of an infinitesimal regular $(n-k)$-polygon, the $(n-k)$-roots of the unit multiplied by an infinitesimal factor times $e^{i\frac{\pi}{n-k}}$. Of course, the perturbation produced by the $\xi_0^\perp(\vec{x};n,k)$ zero modes gives rise to a phase change in the new positions shifted by $-\frac{n}{2}\pi$:
$\theta^\perp(n,k,j)=\theta(n,k,j)-\frac{\pi}{2(n-k)}$.

Although the locations of the zeroes of the Higgs field of a self-dual $n$-vortex obtained as a zero mode fluctuation of a rotationally
symmetric vortex capture the essential features of these stringy objects, it is also interesting to look at the associated magnetic field.
The magnetic fields of a rotationally symmetric vortex perturbed along the $\xi_0(\vec{x};n,k)$ and $\xi_0^\perp(\vec{x};n,k)$ zero modes are:
\begin{eqnarray}
\widetilde{F}_{12}(\vec{x};n,k) &=& F_{12}(\vec{x};n)+\epsilon f_{12}(\vec{x};n,k)=\frac{n}{r}\frac{d\beta_n}{dr}(r)+\epsilon \cos[(n-k)\theta]\cdot r^{n-k-1}\frac{d h_{nk}}{dr}(r)
\label{zmmagper} \\ \widetilde{F}_{12}^\perp(\vec{x};n,k)&=&\frac{n}{r}\frac{d\beta_n}{dr}(r)-\epsilon \sin[(n-k)\theta]\cdot r^{n-k-1}\frac{d h_{nk}}{dr}(r)
\label{zmmagperp}\quad .
\end{eqnarray}
Because the zero mode fluctuations of rotationally symmetric self-dual vortices are still solutions of the first-order equations, their energy density per unit surface area reads
\begin{equation}
{\cal V}(\vec{x};n,k)=\frac{1}{2}\Big(1-\vert \widetilde{\psi}(\vec{x};n,k)\vert^2\Big)\simeq \frac{1}{2}\left(1-f_n^2(r)\right)+\epsilon \cos[(n-k)\theta]r^{n-k-1}\frac{dh_{nk}}{dr}(r) \label{stden} \, \, ,
\end{equation}
which is exactly the magnetic field of the deformed solution, as it should be the case. We emphasize that the information contained in formulas (\ref{psdhf})-(\ref{zmmagperp})
describes the neutral equilibrium fluctuations of any BPS cylindrically symmetric $n$-vortex not merely as the free motion of the Higgs zeroes, the vortex centers, but also tells us how the scalar, vector and magnetic fields are deformed in these processes in a manner as precise as permitted by the limited analytical knowledge of the BPS multivortex solutions. It is clear from (\ref{zeromode4}) that the construction of the $2n$ zero modes of fluctuations of a rotationally symmetric self-dual $n$-vortex is limited to the numerical computation of the $k=0,1,\dots,n-1$ radial form factors $h_{nk}(r)$. In the previous subsection, analytical expressions of these functions were found near $r=0$ and very far from the origin.

It remains to describe the vortex zero modes in the intermediate region by interpolating between these two regimes. Thus, we search for numerically generated solutions $h_{nk}(r)$ using the same procedure as that previously employed to find the self-dual vortex profiles $f_n(r)$ and $\beta_n(r)$. Taking into account that (\ref{ode5}) is a homogeneous linear differential equation, we fix the value of the function at some determined point $r=r_0$, i.e., $h_{nk}(r_0)=c_1$ and allow the value of its derivative at this point $h_{nk}'(r_0)$ to vary. Using a shooting numerical technique, we tune the value of $h_{nk}'(r_0)$ to obtain the exponentially decaying solution, which defines $h_{nk}(r)$ as a normalizable function, otherwise the solution goes to infinity. This procedure allows us to construct $h_{nk}(r)$ in the $[r_0,\infty)$ range. Starting from the same integration constants and the same scheme with a negative step, we obtain a numerical profile of the function $h_{nk}(r)$ also in the $[0,r_0]$ range. The numerical results confirm the theoretical behavior for small and large values of $r$ derived analytically.

We shall now offer graphic representations of the self-dual vortex zero modes $\xi_0(\vec{x};n,k)$ and the corresponding perturbed fields $\widetilde{\psi}$ and $\widetilde{V}$ defined in (\ref{perturbed}) or (\ref{psdhf}) with the aim of gaining some insight into the meaning of the new solutions. We warn the reader that in order to reach an appreciable visualization we shall choose a finite $\epsilon$ instead of taking some infinitesimal value. Graphics describing the 5-vortex zero modes $\xi_0(\vec{x};n,k)$ for every value of the angular momentum $k=0,1,2,3,4$ are collected in Figure 1, which arranges the relevant plots in a table format in order of decreasing value of $k$. Vector plots of the scalar and vector field zero mode fluctuations $\varphi(\vec{x})$ and $a(\vec{x})$ are displayed respectively in the first and third columns of Figure 1. In the second and fourth columns of Figure 1, however, we insert respectively vector plots for the scalar field $\widetilde{\psi}$ and vector potential $\widetilde{V}$ of the perturbed vortex solutions together with a density graphics of the value of its moduli. Here the darker the color is the less the value of the modulus is in such a way that the single vortex centers emerge as shadowed regions in the plots. The first row in Figure 1 illustrates the profile of the zero mode $\xi_0(\vec{x},5,4)$ of angular momentum $k=4$ and the behavior of the corresponding perturbed fields. Under this zero mode fluctuation one of the zeroes of the $n=5$-vortex scalar field $\psi(\vec{x})$ moves along the $x_1$-axes, i.e., one of the five single quanta forming the BPS 5-vortex configuration segregates from the other ones, which remain located at the origin of the plane.

\vspace{0.2cm}

\begin{figure}[H]
\begin{tabular}{c|cccc}
 & 1. $\varphi(\vec{x})$ & 2. $\psi(\vec{x})+\epsilon \, \varphi(\vec{x})$ & 3. $a(\vec{x})$ &  4. $V(\vec{x}) +\epsilon \,  a(\vec{x})$ \\ \hline
\rotatebox{90}{\hspace{0.8cm} 1. $\xi_0(\vec{x},5,4)$} &\includegraphics[width=3.5cm]{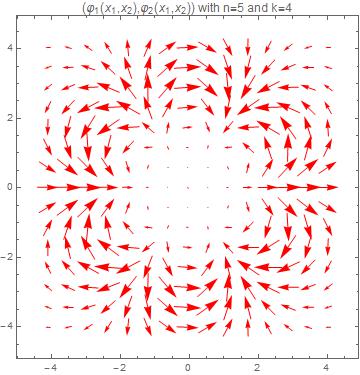} & \includegraphics[width=3.5cm]{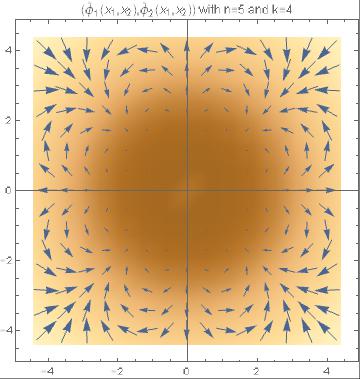} & \includegraphics[width=3.5cm]{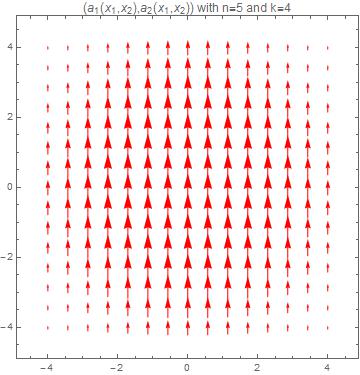} & \includegraphics[width=3.5cm]{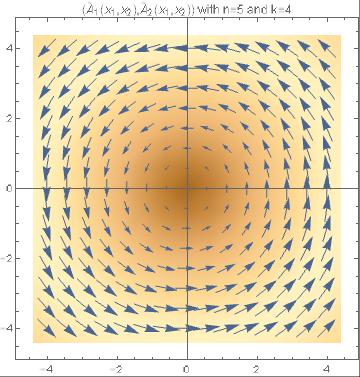} \\
\rotatebox{90}{\hspace{0.8cm} 2. $\xi_0(\vec{x},5,3)$} &\includegraphics[width=3.5cm]{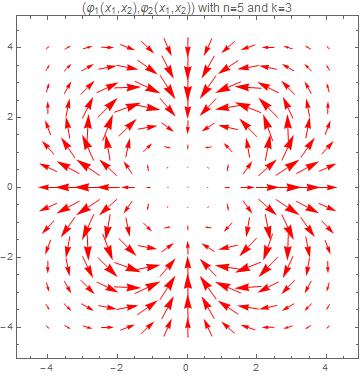} & \includegraphics[width=3.5cm]{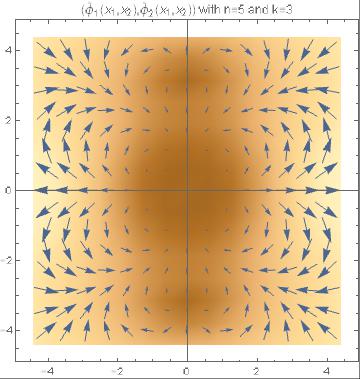} & \includegraphics[width=3.5cm]{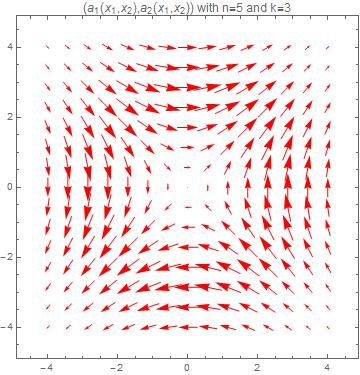} & \includegraphics[width=3.5cm]{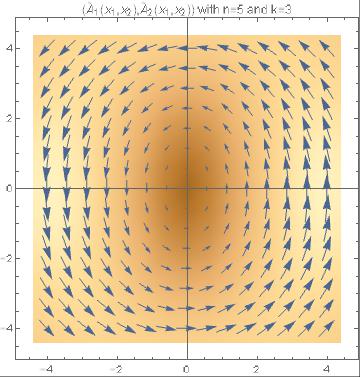} \\
\rotatebox{90}{\hspace{0.8cm} 3. $\xi_0(\vec{x},5,2)$} &\includegraphics[width=3.5cm]{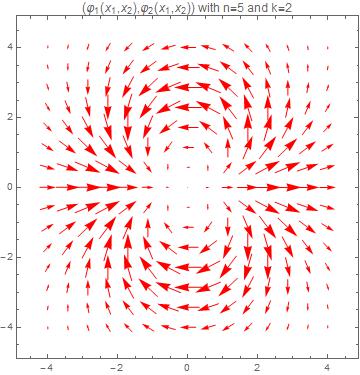} & \includegraphics[width=3.5cm]{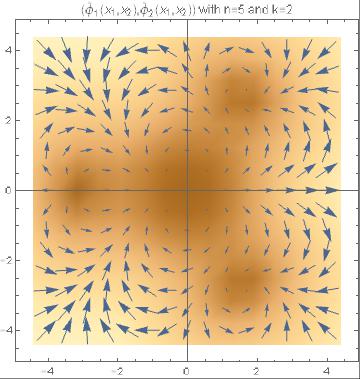} & \includegraphics[width=3.5cm]{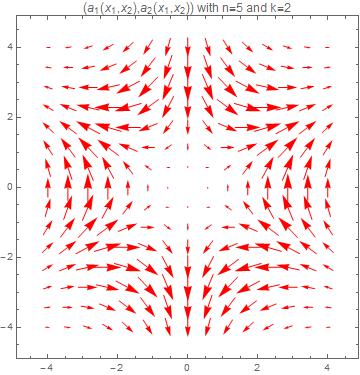} & \includegraphics[width=3.5cm]{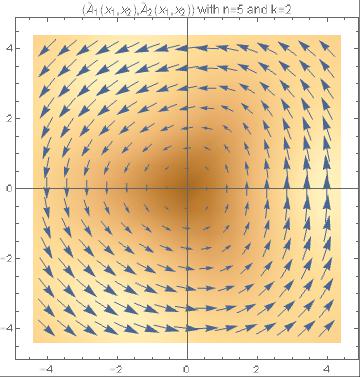} \\
\rotatebox{90}{\hspace{0.8cm} 4. $\xi_0(\vec{x},5,1)$} &\includegraphics[width=3.5cm]{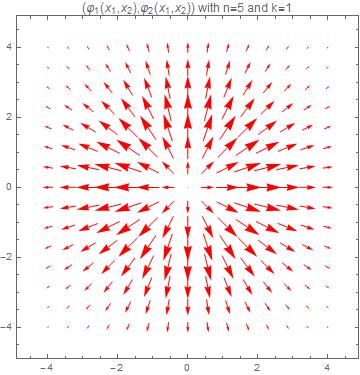} & \includegraphics[width=3.5cm]{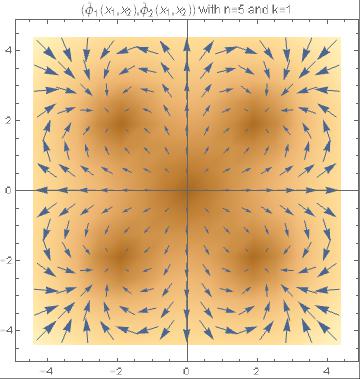} & \includegraphics[width=3.5cm]{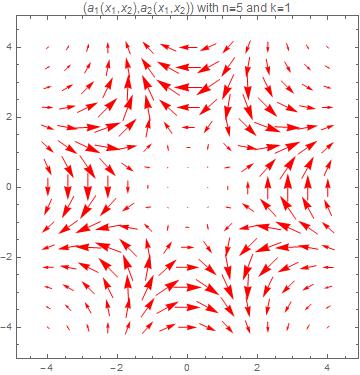} & \includegraphics[width=3.5cm]{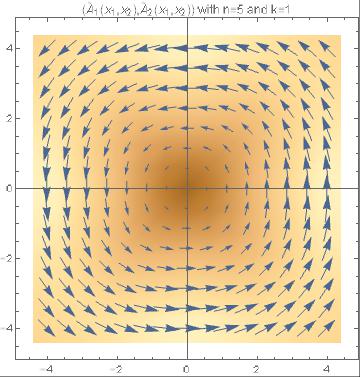} \\
\rotatebox{90}{\hspace{0.8cm} 5. $\xi_0(\vec{x},5,0)$} &\includegraphics[width=3.5cm]{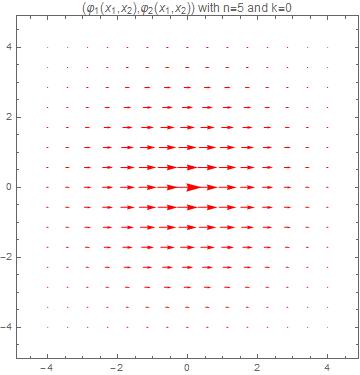} & \includegraphics[width=3.5cm]{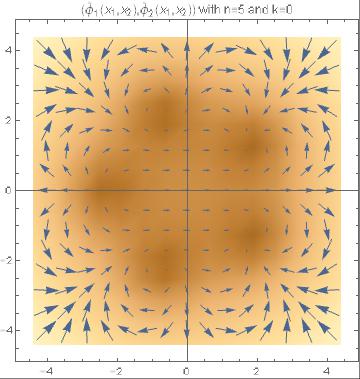} & \includegraphics[width=3.5cm]{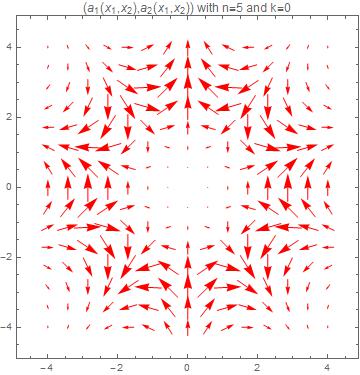} & \includegraphics[width=3.5cm]{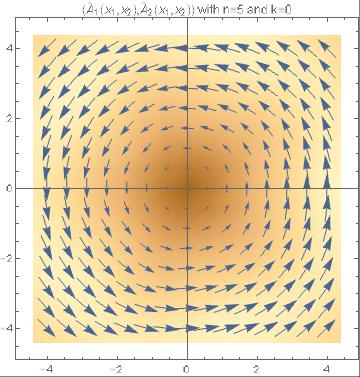} \\
\end{tabular}
\caption{$\xi_0(\vec{x};5,k)$ zero modes of fluctuation around a $5$-vortex and the self-dual $5$-vortex perturbed by the zero mode $\xi_0(\vec{x}; 5,k)$ for the values $k=4,3,2,1,0$.}
\end{figure}

\noindent The corresponding perturbed vector potential $\widetilde{V}(\vec{x})$ is shown in the last Figure plotted in this row. The differences between the perturbed and non-perturbed vector potentials are not so prominent as for the companion scalar fields. In the second row, Figure 1, the zero mode fluctuation of angular momentum $k=3$, $\xi_0(\vec{x},5,3)$, and its influence on the BPS 5-vortex solution is displayed. It is seen in the perturbed scalar field that two of the vortex quanta, initially amalgamated in the 5-vortex center, separate following opposite senses along the $x_2$ direction from the remaining ones which stay at the origin. The fluctuation for angular momentum $k=2$ is illustrated in the third row, Figure 1. Graphical representations of the zero mode $\xi_0(\vec{x},5,2)$ scalar and vector fields  are respectively shown in the first and third columns. Under this zero mode fluctuation of the original 5-vortex configuration only two constituent quanta remain fixed at the origin while the other three magnetic flux quanta are ejected in directions whose relative angles are $\frac{2\pi}{3}$ radians. Next, fourth row, Figure 1, the zero mode $\xi_0(\vec{x},5,1)$ of angular momentum $k=1$ fluctuation fields are plotted. Four quanta of magnetic flux are ejected following the diagonal lines of every quadrant in the plane, while the remaining single quanta stays at the original location. Finally, in the last row and second column in Figure 1, the cylindrically symmetric self-dual vortex solution perturbed by the action of the $k=0$ zero mode scalar field fluctuation is plotted. We observe that all the five single quanta in the original configuration are expelled in the directions determined by the vertices of a regular pentagon.

\section{Boson-vortex bound states. Internal fluctuation modes on BPS cylindrically symmetric vortices}

We shall investigate now the existence of excited fluctuation modes in the pure point spectrum of the ${\cal H}^+$-operator. With respect to the vortex zero modes an important difference is the need of solving second-order differential equations, instead of copying with first-order differential equations systems. To overcome this difficulty we shall take profit from the supersymmetric structure of the variational problem around BPS vortices in the Abelian Higgs model.

\subsection{Particles on BPS topological vortices: translational, internal and scattering modes.}

We first note that the eigenfunctions of ${\cal H}^+$ come in pairs, ${\cal H}^+ \xi^+_\lambda(\vec{x})=\omega_\lambda^2 \xi^+_\lambda(\vec{x})$ and ${\cal H}^+ \xi^{+\perp}_\lambda(\vec{x})=\omega_\lambda^2 \xi^{+\perp}_\lambda(\vec{x})$, which are degenerate and orthogonal to each other. The PDE system equivalent to the spectral condition exhibits also the discrete symmetry arising in the analysis of zero modes even if $\omega^2_\lambda >0$. This symmetry prompts the following Lemma:

\vspace{0.2cm}

\noindent \textsc{Lemma:} Let $\xi^+_\lambda(\vec{x})=(a_r,a_\theta,\varphi_1,\varphi_2)^t$  be an eigenfunction of the second-order  vortex fluctuation operator ${\cal H}^+$ with eigenvalue $\omega_\lambda^2$. Then, $\xi_\lambda^{+\perp}(\vec{x})=(a_\theta, -a_r,\varphi_2,-\varphi_1)^t$ is another eigenfunction of ${\cal H}^+$ with the same eigenvalue $\omega_\lambda^2$, which is orthogonal to, henceforth linearly independent of, $\xi^+_\lambda(\vec{x})$.

\vspace{0.1cm}

\noindent \textsc{Proof:} The spectral condition ${\cal H}^+ \xi_\lambda(\vec{x})= \omega_\lambda^2 \xi_\lambda (\vec{x})$ is tantamount
to the PDE system
\begin{eqnarray}
(-\nabla^2 + |\psi|^2)a_1 -2 \nabla_1 \psi_2 \, \varphi_1 + 2 \nabla_1 \psi_1 \, \varphi_2 &=& \omega_\lambda^2 \,a_1 \nonumber\\
(-\nabla^2 + |\psi|^2)a_2 -2 \nabla_2 \psi_2 \, \varphi_2 + 2 \nabla_2 \psi_1 \, \varphi_2 &=& \omega_\lambda^2 \, a_2 \nonumber\\
-2\nabla_1 \psi_2 \, a_1 - 2 \nabla_2 \psi_2 \, a_2 + \left[-\nabla^2 + {\textstyle\frac{1}{2}} (3|\psi|^2 +2 V_kV_k -1) \right] \varphi_1 -2 V_k \partial_k \varphi_2 &=& \omega_\lambda^2 \,\varphi_1 \label{epdes}\\ 2\nabla_1 \psi_1 \, a_1 + 2 \nabla_2 \psi_1 \, a_2 +2 V_k \partial_k \varphi_1+ \left[-\nabla^2 + {\textstyle\frac{1}{2}} (3|\psi|^2 +2 V_kV_k -1) \right] \varphi_2 &=& \omega_\lambda^2 \, \varphi_2 \nonumber
\end{eqnarray}
by definition. One easily checks that the spectral condition  ${\cal H}^+\xi_n^{+\perp}(\vec{x})=\omega_\lambda^2\xi_n^{+\perp}(\vec{x})$ is also equivalent to the PDE system (\ref{epdes}), provided that $\nabla_2 \psi_2=\nabla_1 \psi_1$ and $\nabla_2\psi_1=-\nabla_1 \psi_2$ hold. These relations are satisfied by the self-dual (BPS) vortices, i.e., the PDE system (\ref{epdes}) is invariant under the
transformation $\varphi_1 \rightarrow \varphi_2$, $\varphi_2 \rightarrow -\varphi_1$, $a_r \rightarrow a_\theta$ and $a_\theta \longrightarrow - a_r$. Therefore, $\xi_\lambda^{+\perp}(\vec{x})$ is an eigenfunction of ${\cal H}^+$ with eigenvalue $\omega^2_\lambda$ if ${\cal H}^+\xi_\lambda^{+}(\vec{x})=\omega_\lambda^2\xi_\lambda^{+}(\vec{x})$. Moreover, $\xi_\lambda^{+\perp}(\vec{x})$ and $\xi_\lambda^{+}(\vec{x})$ are clearly mutually orthogonal. All this proves the Lemma. $\Box$

\vspace{0.2cm}

\noindent Searching for positive eigenfunctions $\xi_\lambda^+(\vec{x})$ of the operator ${\cal H}^+$, we shall impose the background gauge condition in order to eliminate spurious gauge fluctuations.

\vspace{0.2cm}

\noindent \textsc{Proposition:}  Besides the $2 n$ zero modes, the spectrum of the operator ${\cal H}^+$ governing the fluctuations of a rotationally symmetric BPS $n$-vortex also includes excited (positive) orthogonal eigen-modes of two generic classes:

\begin{enumerate}

\item \underline{Class A}. There is a class of fluctuations belonging to the strictly positive spectrum of ${\cal H}^+$ of the form:
\begin{eqnarray}
\xi^{{\rm A}+}_\lambda(\vec{x},n,k)&=& \left( \begin{array}{c} \sin \theta \cos (k\theta) \frac{\partial v_{nk}(r)}{\partial r} - \frac{k}{r} \, v_{nk}(r) \cos \theta \sin(k\theta) \\ -\cos \theta \cos (k\theta) \frac{\partial v_{nk}(r)}{\partial r} - \frac{k}{r} \, v_{nk}(r) \, \sin \theta \sin(k\theta) \\ f_n(r) \, v_{nk}(r)\, \cos(n\theta)\, \cos(k\theta) \\ f_n(r)\, v_{nk}(r)\, \sin(n\theta) \,\cos(k\theta)
 \end{array} \right) \hspace{0.2cm} , \hspace{0.2cm} k=0,1,2,\dots\, , \hspace{0.2cm} \label{excitedmode1}
\end{eqnarray}
\begin{eqnarray}
\chi^{{\rm A}+}_\lambda(\vec{x},n,k)&=& \left( \begin{array}{c} \sin \theta \sin (k\theta) \frac{\partial v_{nk}(r)}{\partial r} + \frac{k}{r} \, v_{nk}(r) \cos \theta \cos(k\theta) \\ -\cos \theta \sin (k\theta) \frac{\partial v_{nk}(r)}{\partial r} + \frac{k}{r} \, v_{nk}(r) \, \sin \theta \cos(k\theta) \\ f_n(r) \, v_{nk}(r)\, \cos(n\theta)\, \sin(k\theta) \\ f_n(r)\, v_{nk}(r)\, \sin(n\theta) \,\sin(k\theta)
 \end{array} \right) \hspace{0.2cm}, \hspace{0.2cm} k=1,2,\dots \, .\hspace{0.2cm} \label{excitedmode2}
\end{eqnarray}
Fluctuations of type $\xi^+_\lambda(\vec{x},n,k)$ are the $k$-term of a cosine Fourier series, while $\chi^+_\lambda(\vec{x},n,k)$ arises in a sine Fourier series. Moreover, $\xi^+_\lambda(\vec{x},n,k)$ and $\chi^+_\lambda(\vec{x},n,k)$ are linearly independent. The radial form factor $v_{nk}(r)$ is determined in both cases as a solution of the 1D Sturm-Liouville problem
\begin{equation}
-\frac{d^2 v_{nk}(r)}{d r^2} -\frac{1}{r} \frac{d v_{nk}(r)}{d r} + \Big[ f_n^2(r)-\omega_\lambda^2 + \frac{k^2}{r^2} \Big] v_{nk}(r)=0 \, \, .
\label{ode55}
\end{equation}
There are excited modes of fluctuations of the form (\ref{excitedmode1})-(\ref{excitedmode2}) built from the solutions of the radial equation (\ref{ode55}) that belong to either the continuous or the discrete spectrum of ${\cal H}^+$.

\begin{itemize}

\item \textit{Bound states}: there exists a discrete set of eigenvalues $\omega_j^2$ confined to the open interval $\omega^2_j\in (0,1)$
such that the associated eigenfunctions of the form (\ref{excitedmode1}) and (\ref{excitedmode2}) are normalizable. In order to belong to $L^2(\mathbb{R}^2)\oplus \mathbb{R}^4$ these excited eigenfunctions must comply with the boundary conditions $\frac{d v_{nk}}{dr}(0)=0$ and $\lim_{r\rightarrow \infty} v_{nk}(r)=0$.

\item \textit{Scattering states}: all the eigenvalues $\omega_\lambda^2(p^2)=p^2+1$, where $p$ is a positive real number, $p\in\mathbb{R}^+-\{0\}$, are admissible. The corresponding eigenfluctuations belong to the continuous ${\cal H}^+$-spectrum
if the form factor $v_{nk}(r)$ satisfies scattering boundary conditions. Thus, the continuous spectrum $\{p^2+1\}_{p\in \mathbb{R}^+-\{0\}}$ emerges from the threshold value $1$.
\end{itemize}
\item \underline{Class B}. There are also eigenfunctions of the form
\[
\xi_\lambda^{{\rm B}+}(\vec{x},n,k)= r^{n-k-1} \left( \begin{array}{c}
h_{nk}(r) \sin [(n-k-1)\theta] \\
h_{nk}(r) \cos[(n-k-1)\theta] \\
-\frac{h_{nk}'(r)}{f_n(r)} \cos (k\theta) \\
-\frac{h_{nk}'(r)}{f_n(r)} \sin (k\theta)
\end{array} \right)
\]
\[
(\xi_\lambda^{{\rm B}+})^\perp(\vec{x},n,k)= r^{n-k-1} \left( \begin{array}{c}
h_{nk}(r) \cos[(n-k-1)\theta] \\
-h_{nk}(r) \sin [(n-k-1)\theta] \\
-\frac{h_{nk}'(r)}{f_n(r)} \sin (k\theta)\\
\frac{h_{nk}'(r)}{f_n(r)} \cos (k\theta)
\end{array} \right)
\]
where $k=0,1,\dots , n-1$ and the radial form factor $h_{nk}(r)$ satisfies
\[
rh_{nk}''(r) + [-1-2k+2n \beta_n(r)] h_{nk}'(r) +r [\omega_\lambda^2 -f^2(r)]h_{nk}(r)=0 \,\, .
\]
There are two types, arising for either $\omega^2_\lambda=0$ or $\omega^2_\lambda>1$. \begin{itemize}
\item \textit{Zero modes}: thoroughly described in the previous Section.

\item \textit{Scattering states}: $h_{nk}(r)$ behaves as an scattering function at infinity.
\end{itemize}
Note that in this class of fluctuation regularity
of the wave functions at the origin restricts the angular momentum to be $0 \leq k\leq n-1$.
\end{enumerate}

\noindent \textsc{Proof:} In order to prove the Proposition stated in the previous subsection we shall exploit the SUSY quantum mechanical structure hidden in the
operators ${\cal D}$, ${\cal D}^\dagger$, ${\cal H}^+$, and ${\cal H}^-$. Apart from the zero modes, the operators ${\cal H}^+$ and ${\cal H}^-$ are isospectral. If $\omega^2_\lambda >0$ the SUSY structure
implies that the eigenfunctions of ${\cal H}$ come in pairs and are related through the supercharges, i.e.,
\[
{\cal H}^-\xi^-_\lambda(\vec{x})=\omega^2_\lambda \xi^-_{\lambda}(\vec{x}) \, \, \Rightarrow \, \, {\cal H}^+\, {\cal D}^\dagger\xi^-_\lambda(\vec{x})=\omega^2_\lambda\, {\cal D}^\dagger \xi^-_{\lambda}(\vec{x})\quad .
\]
Thus, $\xi^+_\lambda(\vec{x})= \frac{1}{\omega_\lambda}\,{\cal D}^\dagger \xi^-_{\lambda}(\vec{x})$ and our strategy in the search for positive eigenfunctions of ${\cal H}^+$ will be to solve the spectral problem of ${\cal H}^-$ and apply the ${\cal D}^\dagger$ operator to $\xi^-_\lambda$ to find the eigenfunctions of ${\cal H}^+$. The reasons are twofold: First, the kernel of ${\cal H}^-$ is null. Second, ${\cal H}^-$ is block-diagonal: It consists of two $1\times 1$ blocks, prompting a complete decoupling of the vector field fluctuations, and one $2\times 2$ block arising because of the decoupling of the scalar field fluctuations from the vector field perturbations. In sum, the spectral problem of ${\cal H}^-$ is divided into three classes of positive eigenfunctions, each class associated with one of the diagonal sub-blocks. We shall now investigate these possibilities separately:
\vspace{0.2cm}

\noindent $\bullet$ \underline{Class A}: There exist ${\cal H}^-$-eigenfunctions of the form
\[
\xi_\lambda^{\rm A -}(\vec{x})=\left( \begin{array}{cccc} a_1(\vec{x}) & 0 & 0 & 0 \end{array} \right)^t
\]
such that $a_1(\vec{x})$ solves the PDE:
\begin{equation}
(-\nabla^2 + |\psi|^2)\, a_1(\vec{x})= \omega_\lambda^2 \, a_1(\vec{x})
\label{pdeA}
\end{equation}
The corresponding SUSY partner ${\cal H}^+$-eigenfunction, which shares the eigenvalue $\omega_\lambda^2$ with $\xi^{\rm A-}_\lambda(\vec{x})$, is:
\begin{equation}
\xi^{\rm A+}_\lambda(\vec{x})= \frac{1}{\omega_\lambda } {\cal D}^\dagger \xi_\lambda^{\rm A-}(\vec{x}) = \frac{1}{\omega_\lambda } \left( \begin{array}{cccc} \partial_2 a_1(\vec{x}) & -\partial_1 a_1(\vec{x}) & \psi_1(\vec{x}) a_1(\vec{x}) & \psi_2(\vec{x}) a_1(\vec{x}) \end{array} \right)^t .\label{autofuncion1}
\end{equation}
It is immediate to check that $\xi_\lambda^{\rm A+}(\vec{x})$ satisfies the background gauge:
\begin{equation}
-\partial_1 [\xi_\lambda^{\rm A+}(\vec{x})]_1-\partial_2 [\xi_\lambda^{\rm A+}(\vec{x})]_2 -\psi_2(\vec{x}) [\xi_\lambda^{\rm A+}(\vec{x})]_3 + \psi_1(\vec{x}) [\xi_\lambda^{\rm A+}(\vec{x})]_4=0 \label{backgroundgauge2} \, \, .
\end{equation}
Thus, fluctuations of the form (\ref{autofuncion1}) are admissible eigenfunctions of ${\cal H}^+$ if $a_1(\vec{x})$ solves the PDE (\ref{pdeA}). The fluctuations $\xi_\lambda^{\rm A+\perp}(\vec{x})$
\[
\xi_\lambda^{\rm A+\perp}(\vec{x})= \frac{1}{\omega_\lambda } \left( \begin{array}{cccc} -\partial_1 a_1(\vec{x}) & -\partial_2 a_1(\vec{x}) & \psi_2(\vec{x}) a_1(\vec{x}) & -\psi_1(\vec{x}) a_1(\vec{x}) \end{array} \right)^t
\]
are orthogonal to $\xi_\lambda^{\rm A+}(\vec{x})$ and belong to the spectrum of ${\cal H}^+$, as shown in the previous Lemma. These eigenfunctions, however, do not satisfy the background gauge condition. Equation (\ref{backgroundgauge2}) applied to $\xi_\lambda^{\rm A+\perp}(\vec{x})$ requires $(-\nabla^2 + |\psi|^2)a_1(\vec{x})=0$, which is only solved by $a_1(\vec{x})=0$ and therefore $\xi_\lambda^{\rm A+\perp}(\vec{x})=0$. These modes of fluctuation are not allowed by the background gauge. We are left with the fluctuations $\xi_\lambda^{\rm A+}(\vec{x})$ of the form (\ref{autofuncion1}), which are eigenfunctions of ${\cal H}^+$ if $a_1(\vec{x})$ satisfies the PDE (\ref{pdeA}), i.e.,
\begin{equation}
-\frac{\partial^2 a_1}{\partial r^2} - \frac{1}{r} \frac{\partial a_1}{\partial r} - \frac{1}{r^2} \frac{\partial^2 a_1}{\partial \theta^2} + [f_n^2(r)-\omega_\lambda^2 ] a_1=0 \label{pdeApolar}
\end{equation}
when written in polar coordinates. The separation ansatz
\begin{equation}
a_1(\vec{x})=v_{nk}(r) \cos (k\theta) \quad {\rm or} \quad a_1(\vec{x})=v_{nk}(r) \sin (k\theta) \label{sepvar}
\end{equation}
converts the PDE (\ref{pdeApolar}) into the ODE (\ref{ode55}) for the radial form factor $v_{nk}(r)$ both if the angular dependence is in cosine or in sine. Univalued fluctuations demand that the wave number $k$ must be a positive integer number $k=1,2\dots$ in the sine case and $k=0,1,2\dots$ in the cosine case.

The ODE (\ref{ode55}) is no more than a radial Schr$\ddot{\rm o}$dinger differential equation with a potential well
\[
V_{\rm eff}(r;n,k)=f_n^2(r) + \frac{k^2}{r^2} \, \, \, ,
\]
which includes a centrifugal barrier when $k\neq 0$, is bounded below and tends to a positive constant at infinity: $\lim_{r\rightarrow \infty} V_{\rm eff}(r)=1$. Substitution of the factorized expressions (\ref{sepvar}) into the fluctuations of the form (\ref{autofuncion1}) produces the expressions (\ref{excitedmode1}) and (\ref{excitedmode2}) given in the Proposition for the excited modes of fluctuation $\xi_\lambda^{\rm A+}(\vec{x})$ and $\chi_\lambda^{\rm A+}(\vec{x})$. The norm of the $\xi_\lambda^{\rm A+}(\vec{x})$ eigenfunctions, for example, is simplified by means of their SUSY nature:
\begin{equation}
\|\xi^{\rm A+}_\lambda(\vec{x})\|^2 = \Big\|\frac{1}{\omega_\lambda} {\cal D}^\dagger \xi_\lambda^{\rm A-}(\vec{x})\, \Big\|^2 =   \| \xi_\lambda^{\rm A-}(\vec{x})\, \|^2 =\int_{\mathbb{R}^2} d^2 x \, a_1^2 (\vec{x}) = C_{k} \pi\int_0^\infty r\, v_{nk}^2(r)\, dr \label{normalization4}
\end{equation}
where $C_k=1$ if $k\neq 0$ and $C_k=2$ if $k=0$. Normalizability of $\xi_\lambda^{\rm A+}$ thus requires regularity at the origin, exponential decay at infinity, and a sufficiently smooth interpolation in between, to $v_{nk}(r)$. We shall now investigate these regimes:

\begin{itemize}
\item \textit{Regularity of the funci\'on $v_{nk}$ at the origin:} Near the origin, one solves for $v_{nr}(r)$ by applying the Frobenius method to the ODE (\ref{ode55}) in the vicinity of the regular singular point $r=0$, i.e., one plugs the series expansion $v_{nr}(r)=r^s \sum_{j=0}^\infty v_j^{(n,k)} r^j$ near $r=0$ starting from the condition $v_0^{(n,k)}\neq 0$ into the ODE (\ref{ode55}). This procedure trades the ODE (\ref{ode55}) for the two term recurrence relation
\begin{equation}
\sum_{j=0}^\infty \Big[ -(s+j)^2 + k^2 \Big] v_j^{(n,k)} r^j + \sum_{j=2}^\infty \Big[ - \omega_\lambda^2 v_{j-2}^{(n,k)}  r^j \Big] + {\cal O}(r^{2n+2})=0 \label{recurrencia2}
\end{equation}
where the coefficients $v_j^{(n,k)}$ are grouped by powers. Annihilation of the $j=0$-term induces the indicial equation $-s^2+k^2=0$. The characteristic exponents are $s=\pm k$ but we shall choose $s=k$ because our solutions are such that $k\geq 0$. The recurrence relations between the odd coefficients arising from (\ref{recurrencia2}) means that these coefficients vanish at least up to the $2n+1$ coefficient. The recurrence relations (\ref{recurrencia2}) between even coefficients $j=2 i, i=1,2, \cdots$ imply that $-4 i (i+k)v_{2i}^{(n,k)} =\omega_\lambda^2 v_{2i-2}^{(n,k)}$ for $i=1,2,\dots$. Therefore, near the origin we find the solution $v_{nk}(r) \approx v_0^{(n,k)}  r^k + v_2^{(n,k)}  r^{k+2} +\dots$ which is regular and presents no problems in the normalizability of $\|\xi^{\rm A+}_\lambda(\vec{x})\|$.

\item \textit{Asymptotic behavior of the function $v_{nk}(r)$:} The asymptotic behavior of the function $v_{nk}(r)$ near infinity is governed by the differential equation
\[
-\frac{d^2 v_{nk}(r)}{d r^2} - \frac{1}{r} \frac{d v_{nk}(r)}{d r} + \Big[1-\omega_\lambda^2 + \frac{k^2}{r^2} \Big]v_{nk}(r) =0 \, \, ,
\]
which is a Bessel or modified Bessel equation. Thus,
\[
v_{nk}(r) \stackrel{r\rightarrow \infty}{\longrightarrow} \left\{ \begin{array}{l} \overline{C}_1\, I_k \Big(\sqrt{1-\omega_\lambda^2}\, r\Big) + \overline{C}_2 \, K_k \Big(\sqrt{1-\omega_\lambda^2} \, r\Big) \hspace{0.9cm} , \hspace{0.9cm} \mbox{if} \quad \omega_\lambda \in (0,1) \\  \overline{C}_1 \, J_k \Big(-\sqrt{\omega_\lambda^2-1} \, r \Big) + \overline{C}_2 \, Y_k \Big(-\sqrt{\omega_\lambda^2-1} \, r \Big) \hspace{0.5cm} , \hspace{0.5cm} \mbox{if} \quad \omega_\lambda \in [1,\infty)   \end{array} \right.
\]
where $I_k$, $K_k$, $J_k$ and $Y_k$ belong to the Bessel function catalogue. The asymptotic behaviour of form factor $v_{nk}(r)$ is accordingly
{\small\[
v_{nk}(r) \stackrel{r\rightarrow \infty}{\longrightarrow} \left\{ \begin{array}{l} \frac{\overline{C}_1}{\sqrt{r}}\, {\rm exp} \Big[\sqrt{1-\omega_\lambda^2}\, r\Big] + \frac{\overline{C}_2}{\sqrt{r}}  \,{\rm exp} \Big[\sqrt{1-\omega_\lambda^2} \, r\Big] \hspace{2.6cm}  \hspace{2.6cm} \mbox{if} \quad \omega_\lambda \in (0,1)  \\ \frac{\overline{C}_1}{\sqrt{r}} \, \cos \Big[\sqrt{\omega_\lambda^2-1} \, r +(k+\frac{1}{2})\frac{\pi}{2}\Big] + \frac{\overline{C}_2}{\sqrt{r}} \, \sin \Big[-\sqrt{\omega_\lambda^2-1} \, r - (k+\frac{1}{2})\frac{\pi}{2}\Big] \quad  \quad  \quad \mbox{if} \quad \omega_\lambda \in [1,\infty)  \end{array} \right.
\]}
The genuine Bessel function $J_k$ and $Y_k$ exhibiting a non-normalizable oscillatory asymptotic behaviour are solutions of the above ODE when $\omega_\lambda^2 >1$. A continuous spectrum arises in the  $\omega_\lambda\in [1,\infty)$ range, i.e., for energies above the scattering threshold $\omega^2_\lambda=1$. Below this threshold, in the $\omega_\lambda^2 \in (0,1)$ range, boson-vortex bound states may exist because of the exponential asymptotic behavior of the modified Bessel functions $I_k$ and $K_k$. To find the eigenvalues in the point spectrum of both ${\cal H}^\pm$ it is necessary to identify a particular solution whose behaviour near the origin is regular as indicated previously and exhibit a decreasing exponential tail ($\overline{C}_1=0$ in the asymptotic solutions given before). The difficulty in finding this type of bound state fluctuations is the identification of the discrete set of $\omega_j^2\in(0,1)$ enabling their existence. In sum, in the discrete spectrum of ${\cal H}^\pm$ we expect to find SUSY pairs of positive modes of fluctuation of the generic form (\ref{autofuncion1}) belonging to the $L^2(\mathbb{R}^2)\oplus \mathbb{R}^4$ Hilbert space.
\end{itemize}

\noindent Note that the second choice in our strategy of studying the decoupled blocks in ${\cal H}^-$, i.e., the search for ${\cal H}^-$-eigenfunctions of the form
\[
\zeta_\lambda^{\rm A-}(\vec{x})=\left( \begin{array}{cccc} 0 & a_2(\vec{x}) & 0 & 0 \end{array} \right)^t
\]
where $a_2(\vec{x})$ must be a solution of the spectral ODE $(-\nabla^2 + |\psi|^2) \, a_2(\vec{x})= \omega_n^2 \, a_2(\vec{x})$, has already been discussed because of the discrete symmetry shown in the last Lemma. This symmetry connects the eigenfunctions $\xi_\lambda^{A-}(\vec{x})$ and $\zeta_\lambda^{A-\perp}(\vec{x})$ and consequently its SUSY partners. In particular, the SUSY partner ${\cal H}^+$-eigenfunction adopt the form
\[
\zeta^{\rm A+}_\lambda(\vec{x})= \frac{1}{\omega_\lambda} {\cal D}^\dagger \zeta_\lambda^{\rm A-} (\vec{x})= \frac{1}{\omega_\lambda} \left( \begin{array}{cccc} \partial_1 a_2(\vec{x}) & \partial_2 a_2(\vec{x}) & -\psi_2(\vec{x}) a_2(\vec{x}) & \psi_1(\vec{x}) a_2(\vec{x}) \end{array} \right)^t
\]
such that the background gauge condition (\ref{backgroundgauge2}) becomes the PDE: $(-\nabla^2 + |\psi|^2)a_2(\vec{x})=0$. The unique regular solution is $a_2(\vec{x})=0$ and the background gauge discards this type of fluctuation mode. The  perpendicular eigenfunction $\zeta^{\rm A+\perp}_\lambda(\vec{x})$, however,
\begin{equation}
\zeta_\lambda^{\rm A+\perp}(\vec{x})= \frac{1}{\omega_\lambda} {\cal D}^\dagger \zeta_\lambda^{\rm A-\perp} (\vec{x})=  \frac{1}{\omega_\lambda} \left( \begin{array}{cccc} \partial_2 a_2(\vec{x}) & -\partial_1 a_2(\vec{x}) & \psi_1(\vec{x}) a_2(\vec{x}) & \psi_2(\vec{x}) a_2(\vec{x}) \end{array} \right)^t \label{autofuncion2}
\end{equation}
automatically satisfies the background condition (\ref{backgroundgauge2}). But by comparing (\ref{autofuncion1}) and (\ref{autofuncion2}) we conclude that the result found encompasses exactly the same type of eigenfluctuations of ${\cal H}^+$ as those described before because $a_1(\vec{x})$ and $a_2(\vec{x})$ are both solutions of the same equation (\ref{pdeA}).

\vspace{0.2cm}

\noindent $\bullet$ \underline{Class B}. The lower $2\times 2$  block-diagonal matrix inside ${\cal H}^-$ acts only on the scalar field fluctuations. There exist
another class of fluctuations of the form
\begin{equation}
\xi_\lambda^{\rm B-}(\vec{x})=\left( \begin{array}{cccc} 0 & 0 & \varphi_1(\vec{x}) & \varphi_2(\vec{x}) \end{array} \right)^t \label{eigenB}
\end{equation}
which are a priori candidates to belong to the spectrum of ${\cal H}^-$, paired with its SUSY partners in the spectrum of ${\cal H}^+$. The scalar fluctuations must be solutions of the ${\cal H}^-$-spectral
PDE system:
\begin{equation}
\left( \begin{array}{cc} -\nabla^2 + {\textstyle \frac{1}{2}} \, ( |\psi|^2+1) + V_k V_k & - 2 \,V_k \,\partial_k \\
2\, V_k \,\partial_k & -\nabla^2 + {\textstyle \frac{1}{2}} \, ( |\psi|^2+1) + V_k V_k \end{array} \right)
\left( \begin{array}{c} \varphi_1(\vec{x}) \\ \varphi_2 (\vec{x}) \end{array} \right) =  \omega_\lambda^2 \left( \begin{array}{c} \varphi_1 (\vec{x}) \\ \varphi_2(\vec{x}) \end{array} \right) \label{odeC}
\end{equation}
Assuming that eigenfunctions of ${\cal H}^-$ are obtained by solving (\ref{odeC}), the corresponding SUSY partner ${\cal H}^+$-eigenfunctions are:
\begin{equation}
\xi^{\rm B+}_\lambda (\vec{x})= \frac{1}{\omega_\lambda} {\cal D}^\dagger \xi_\lambda^{\rm B-}(\vec{x})=  \frac{1}{\omega_\lambda} \left( \begin{array}{c}
\psi_1(\vec{x}) \, \varphi_1(\vec{x}) + \psi_2(\vec{x}) \, \varphi_2(\vec{x}) \\
-\psi_2 (\vec{x})\, \varphi_1 (\vec{x}) + \psi_1 (\vec{x})\, \varphi_2(\vec{x}) \\
\partial_2 \varphi_1(\vec{x}) -\partial_1 \varphi_2(\vec{x}) + V_1(\vec{x}) \varphi_1(\vec{x}) + V_2(\vec{x}) \varphi_2(\vec{x})\\
\partial_1 \varphi_1(\vec{x}) + \partial_2 \varphi_2 (\vec{x}) - V_2(\vec{x})\varphi_1(\vec{x}) + V_1 (\vec{x}) \varphi_2(\vec{x})
 \end{array} \right) \label{autofuncion3} \, \, ,
\end{equation}
\begin{equation}
\xi_\lambda^{\rm B+\perp}(\vec{x})=\frac{1}{\omega_\lambda} {\cal D}^\dagger \xi_\lambda^{\rm B-\perp
}(\vec{x})= \frac{1}{\omega_\lambda} \left( \begin{array}{c}
-\psi_2 (\vec{x}) \, \varphi_1(\vec{x}) + \psi_1 (\vec{x}) \, \varphi_2(\vec{x}) \\
-\psi_1 (\vec{x}) \, \varphi_1(\vec{x}) -\psi_2 (\vec{x}) \, \varphi_2 (\vec{x}) \\
\partial_1 \varphi_1 (\vec{x}) + \partial_2 \varphi_2 (\vec{x}) -V_2(\vec{x}) \varphi_1 (\vec{x}) + V_1 (\vec{x}) \varphi_2(\vec{x}) \\ -\partial_2 \varphi_1(\vec{x}) +\partial_1 \varphi_2 (\vec{x}) - V_1(\vec{x}) \varphi_1 (\vec{x}) -V_2(\vec{x}) \varphi_2(\vec{x})
 \end{array} \right)\label{autofuncion4} \, \, .
\end{equation}
It is easy to check that the fluctuations (\ref{autofuncion3}) and (\ref{autofuncion4}) verify the background gauge (\ref{backgroundgauge2}) provided that $\nabla_2 \psi_2=\nabla_1 \psi_1$ and $\nabla_2\psi_1=-\nabla_1\psi_2$ hold. But these first-order PDE are obeyed by the self-dual vortices, and hence $\xi_\lambda^{\rm B+}(\vec{x})$ and $\xi_\lambda^{\rm B+\perp}(\vec{x})$ are admissible as candidates to be ${\cal H}^+$-eigenfunctions. Now we shall demonstrate the lack of discrete eigenfunctions of the form (\ref{eigenB}) in the spectrum of ${\cal H}^-$ and consequently in the spectrum of ${\cal H}^+$. We shall also analyze the connection of this type of eigenmodes with the zero modes:

\begin{itemize}
\item \textit{Lack of discrete spectrum of ${\cal H}^-$ and ${\cal H}^+$:} It is easy to check that the $2 \times 2$ block inside ${\cal H}^-$ acting on the scalar fluctuations
\begin{equation}
{\cal H}^-|_{(\varphi_1,\varphi_2)} = \left( \begin{array}{cc} -\nabla^2 + {\textstyle \frac{1}{2}} \, ( |\psi|^2+1) + V_k V_k & - 2 \,V_k \,\partial_k \\
2\, V_k \,\partial_k & -\nabla^2 + {\textstyle \frac{1}{2}} \, ( |\psi|^2+1) + V_k V_k \end{array} \right) \label{operator2}
\end{equation}
can be factorized as
\begin{equation}
{\cal H}^-|_{(\varphi_1,\varphi_2)} = {\cal E}^\dagger {\cal E} +1 \label{factorization03}
\end{equation}
in terms of the first order partial differential operator
\[
{\cal E}= \left( \begin{array}{cc} \partial_1 + V_2(\vec{x}) & -\partial_2+V_1(\vec{x}) \\ \partial_2-V_1(\vec{x}) & \partial_1 + V_2(\vec{x}) \end{array} \right)
\]
and its adjoint. The factorization (\ref{factorization03}) means that the eigenvalues $\omega_\lambda^2$ of the spectral problem (\ref{odeC}) are greater than or equal to the threshold value 1. Thus, there are no positive ${\cal H}^-$-bound states of this class and only a continuous spectrum emerges. Nevertheless, the nature of eigenfunctions of eigenvalue exactly equal to the scattering threshold $\omega_1^2=1$ is unclear at this
point and demands a closer investigation.

The separation in the polar coordinates ansatz
\begin{equation}
\varphi_1(\vec{x}) = - \, u_{nk}(r) \sin[(k+1)\theta] \hspace{0.5cm},\hspace{0.5cm} \varphi_2(\vec{x})=u_{nk}(r) \cos[(k+1)\theta] \label{ansatz1}
\end{equation}
converts the PDE system (\ref{odeC}) into the spectral ODE:
\begin{equation}
-\frac{d^2 u_{nk}(r)}{d r^2} - \frac{1}{r} \frac{d u_{nk}}{d r} + \Big[ \frac{1}{2} (1+f_n^2(r)) + \frac{(n \beta_n(r)-(1+k))^2}{r^2} -\omega_\lambda^2 \Big]u_{nk}(r) =0 \label{odeC3}
\end{equation}
with the radial form factor $u_{nk}(r)$ as unknown. This is a radial Schr$\ddot{\rm o}$dinger differential equation with an effective potential well:
\[
V_{\rm eff}^{\rm B}(r)=\frac{1}{2} [1+f_n^2(r)] + \frac{1}{r^2}[n \beta_n(r)-(1+k)]^2 \quad .
\]
The asymptotics of $V^{\rm B}_{\rm eff}$ respectively near $r=0$ or infinitely far apart of the vortex core $r=\infty$ are:
\[
V^{\rm B}_{\rm eff}(r)\stackrel{r\to 0}{\simeq}\frac{1}{2}\left(1+d_n r^{2n}\right)+\frac{1}{r^2}\left(n \, e_2\, r^2-(1+k)\right)^2 \quad , \quad \lim_{r\rightarrow \infty} V^{\rm B}_{\rm eff}(r)=1 \, \, .
\]
which reinforces the previous claim about the emergence of the continuous spectrum on the threshold value $\omega_\lambda^2=1$. Indeed the $\omega_\lambda^2=1$  $u_{nk}^{\{1\}}(r)$-eigenfunction belong to the kernel of the second-order differential operator ${\cal R}$
\[
{\cal R}_{\{1\}}\, u_{nk}^{\{1\}}(r)= \Big[-\frac{d^2}{d r^2} - \frac{1}{r} \frac{d }{d r} + \frac{1}{2}(f_n^2(r)-1) + \frac{(n\beta_n(r)-(1+k))^2}{r^2} \Big] u_{nk}^{\{1\}}(r) =0 \, \, ,
\]
a relation immediately derived from the ODE (\ref{odeC3}) with $\omega_\lambda^2=1$. The operator ${\cal R}_{\{1\}}$ admits a (Infeld-Hull) factorization in the form ${\cal R}_{\{1\}}= {\cal L}^\dagger {\cal L}$, where ${\cal L}$ and ${\cal L}^\dagger$ are the first-order differential operators
\[
{\cal L}=-\frac{d}{d r} + \frac{W_{nk}(r)}{r} \hspace{0.5cm},\hspace{0.5cm}
{\cal L}^\dagger =\frac{d}{d r} +\frac{1}{r}+\frac{W_{nk}(r)}{r} \hspace{1cm}; \hspace{1cm} W_{nk}(r)=1+k-n\beta_n(r) \quad .
\]
It is thus clear that the eigenfunctions with $\omega^2=1$ correspond to the $u_{nk}^{\{1\}}(r)$-eigenfunctions belonging to the kernel of ${\cal L}$: ${\cal L} u_{nk}^{\{1\}}(r)=[-\frac{d}{dr}+\frac{W_{nk}}{r}(r)]u_{nk}^{\{1\}}(r)=0$. The radial form factors of these \lq\lq one\rq\rq{} modes of the ${\cal H}^+$ operator
are easily integrated:
\[
u_{nk}^{\{1\}}(r)\propto {\rm exp} \Big[\int \, dr \, \frac{W_{nk}}{r}(r) \Big] = r^{k+1} e^{-n \int \frac{\beta_n(r)}{r}dr} \, \, .
\]
The zero at the origin is needed to compensate the centrifugal barrier. Thus, $u_{nk}^{\{1\}}(r)$ lacks physical nodes in $\mathbb{R}^+$, denoting the absence of lower eigenfunctions. This means that there are no bound states in the discrete spectrum of ${\cal H}^\pm$ within this class B of fluctuations that only includes the zero modes in the point spectrum of ${\cal H}^+$. A closer look at the behaviour of the  $u_{nk}^{\{1\}}(r)$ one modes close to infinity $ u_{nk}^{\{1\}}(r)\simeq_{r\to \infty}r^{k+1-n}$
reveals interesting features of these radial wave functions living at the threshold of the continuous spectrum of ${\cal H}^-$ and ${\cal H}^+$. If $n=k+1$, $u_{n,n-1}^{\{1\}}(r)$ tends to a constant at infinity and go to zero at the origin as $r^n$. These  modes are thus akin to the half-bound states arising in the fluctuations of sine-Gordon and $\phi^4$ 1D kinks, respectively $\tanh x$ and $3 \tanh^2 x-1$. Note that the half-bound states tend respectively to $\pm 1$ or $2$ at $x=\pm\infty$ and have either one node at $x=0$ or two nodes at $x=\pm {\rm arctan}\frac{1}{\sqrt{3}}$, see e.g. Reference \cite{Alonsoiz}. This type of modes with $k=0,1,2, \cdots , n-2$ decays to zero at
infinity like negative powers of $r$, at most as $r^{-1}$, and these states could be interpreted as a new type of \lq\lq fractionary\rq\rq bound states by some refinement of the 2D Levinson theorem.

\item \textit{BPS vortex zero modes versus class B eigenfunctions:} Now we shall unveil the connection between the zero modes studied in Section \S.4 and the Class B eigenfunctions discussed in this Section. In order to perform the analysis just mentioned in a form as close as possible to the description of vortex zero modes elaborated in Section \S.4, we express the function $u_{nk}(r)$ appearing in (\ref{odeC3}) in the form: $u_{nk}(r) = \frac{g_{nk}(r)}{f_n(r)}$. The ODE (\ref{odeC3}) becomes
{\small\begin{equation}
g_{nk}''(r)+ \,[1-2n +2n\,\beta_n(r)\,]\, \frac{g_{nk}'}{r}\,(r) -\left[(1+k-n)(1+k+n-2n\, \beta_n(r))+r^2 f_n^2(r)-r^2 \omega_\lambda^2 \right] \, \frac{g_{nk}}{r^2}(r)=0 \label{ode66} \, \, ,
\end{equation}}
which is exactly the ODE (\ref{ode3}) except for the term proportional to the frequency $\omega_\lambda^2$. Therefore, one sees that the positive eigenmodes of clas B in the spectrum of ${\cal H}^+$ belong to the same class as the BPS vortex zero modes.

The asymptotic behavior of the radial form factor $g_{nk}(r)$ is determined by the Bessel ODE
\[
-r^2 g_{nk}''(r)-r g_{nk}'(r) + [(1+k-n)^2 + r^2 (1-\omega_\lambda^2)]g_{nk}(r)=0
\]
such that:
\[
g_{nk}(r) \stackrel{r\rightarrow \infty} {\longrightarrow} C_1  \, J_{1+k-n}\Big(-\sqrt{\omega_\lambda^2-1} \, r\Big)  +  C_2 \, Y_{1+k-n}\Big(-\sqrt{\omega_\lambda^2-1} \, r\Big) \, \, ,
\]
revealing the oscillatory asymptotic behavior of $g_{nk}(r)$, which confirms the existence of only continuous spectrum for $\omega_\lambda^2 \geq 1$.

Application of the Frobenius method to the differential equation (\ref{ode66}) around the regular singular point $r=0$ reveals similar behaviour near the origin of the positive eigenfunctions in this class B to the behaviour of vortex zero modes close to the origin:  $g_{nk}(r)=r^{n-k-1} h_{nk}(r)$, where we also saw that $h_{nk}(r) \approx_{r\to 0} c_0^{(n,k)} + c_{2k+2}^{(n,k)} r^{2k+2} + {\cal O}(r^{2k+3})$ with $c_0^{(n,k)}\neq 0$. The eigenfunctions (\ref{autofuncion3}) and (\ref{autofuncion4}) thus read
\[
\xi_\lambda^{\rm B+}(\vec{x})= r^{\overline{m}} \left( \begin{array}{c}
h_{nk}(r) \sin [\overline{m}\,\theta] \\
h_{nk}(r) \cos[\overline{m}\,\theta] \\
-\frac{h_{nk}'(r)}{f_n(r)} \cos (k\,\theta) \\
-\frac{h_{nk}'(r)}{f_n(r)} \sin (k\,\theta)
\end{array} \right)
\, \, \, , \, \, \, \quad
\xi_\lambda^{\rm B+\perp}(\vec{x})= r^{\overline{m}} \left( \begin{array}{c}
h_{nk}(r) \cos[\overline{m}\,\theta] \\
-h_{nk}(r) \sin [\overline{m}\,\theta] \\
-\frac{h_{nk}'(r)}{f_n(r)} \sin (k\,\theta)\\
\frac{h_{nk}'(r)}{f_n(r)} \cos (k\,\theta)
\end{array} \right)
\]
with $\overline{m}=n-k-1$. In this case the radial form factor $h_{nk}(r)$ satisfies the ODE
\[
r\,h_{nk}''(r) + [-1-2k+2n \beta_n(r)] h_{nk}'(r) +r [\omega_\lambda^2 -f_n^2(r)]h_{nk}(r)=0 \, \, ,
\]
which differs from the vortex zero mode equation only in the $\omega_\lambda^2$-term. Regularity at the origin, however, of the positive eigenfluctuations $\xi_\lambda^{\rm B+}(\vec{x})$ and $\xi_\lambda^{\rm B+\perp}(\vec{x})$ requires that $0\leq k \leq n-1$, as in the zero modes of fluctuation case. This proves the second, class B, proposal in the Proposition.
\end{itemize}

\noindent Finally it remains to prove the orthogonality of the eigen-modes of ${\cal H}^+$. Orthogonality between eigenfunctions belonging to different classes is guaranteed by the conservation of the scalar product in the SUSY partnership:
\[
\left\langle \xi_\lambda^{A+}(\vec{x}) , \xi_{\lambda'}^{B+}(\vec{x}) \right\rangle = \left\langle \xi_\lambda^{A-}(\vec{x}) , \xi_{\lambda'}^{B-}(\vec{x}) \right\rangle =0
\]
because class A and B eigenfunctions of ${\cal H}^-$ are
clearly orthogonal.  Orthogonality between eigenfunctions belonging to the same class with different angular dependence is established
by the Fourier theory and, in general, Sturm-Liouville theory of the radial Schr$\ddot{\rm o}$dinger differential equations guarantees
that eigenfunctions having different eigenvalue are orthogonal. $\Box$

\subsection{Portrait of bosonic bound states on BPS cylindrically symmetric vortices }

In this last Section we shall try to elucidate the existence and to understand the structure of excited fluctuations of class A belonging to the discrete spectrum of ${\cal H}^+$ with positive eigenfunctions lower than $1$. Physically, these fluctuation modes obey to certain combinations of the scalar and vector field fluctuations trapped by the rotationally symmetric BPS vortex. Thus, fluctuations with these properties correspond to boson-vortex bound states and our goal is to identify some of these possible bound states as well as to investigate their properties. This task is more difficult than the search for vortex zero modes because we need to find numerically not only the eigenfunctions but also the discrete eigenvalues $\omega_j^2$ in the $(0,1)$ interval.

\subsubsection{Numerical procedure for finding BPS vortex excited modes of fluctuation of bound state type}

The search for and the analysis of some fluctuations in the discrete spectrum of ${\cal H}^+$ reduces to the numerical computation of the radial form factor $v_{nk}(r)$ in the ODE (\ref{ode55}). The existence of boson-vortex bound states can be physically understood analysing the properties of the effective potential wells
\begin{displaymath}
V_{\rm eff}(\vec{x},n,k)=V(r)+\frac{k^2}{r^2}
\end{displaymath}
which arise in the appropriate Schr$\ddot{\rm o}$dinger equation. These equations can also be understood as governing the quantum planar motion of a particle moving in a central potential. Accordingly, we expect to find bound states when the angular momentum is low enough. Our strategy to achieve this finding is to employ a second-order finite-difference scheme which simulates the differential equation (\ref{ode55}) by the recurrence relations
\begin{equation}
-\frac{v_{nk;j}^{(i+1)} - 2 v_{nk;j}^{(i)} +v_{nk;j}^{(i-1)}}{(\Delta x)^2} - \frac{v_{nk;j}^{(i+1)} - v_{nk;j}^{(i-1)}}{2i (\Delta x)^2} + \Big[ f_n^2(i \Delta x) + \frac{k^2}{i^2 (\Delta x)^2} \Big] v_{nk;j}^{(i)} = \omega_{nk;j}^2 \, v_{nk;j}^{(i)} \label{erec1}
\end{equation}
where we have confined the problem to the interval $[0,r_{\rm max}]$ for a large $r_{\rm max}$ enough. We denote $v_{nk;j}^{(i)}=v_{nk;j}(i\Delta x)$, with $\Delta x= {r_{\rm max}}/{N}$, and choose a mesh of $N$ points with $i=0,1,2, \cdots , N$. The eigenfunction and the eigenvalue depend on the values of the angular momentum $k$ and the quantized magnetic flux $n$. The index $j$ labels the discrete eigenfunctions. The contour conditions are:
\[
(1) \, \, -\frac{4}{3} \frac{v_{nk;j}^{(2)}-v_{nk;j}^{(1)}}{(\Delta x)^2} + \Big[f_n^2 (\Delta x) + \frac{k^2}{(\Delta x)^2} \Big] v_{nk;j}^{(1)} = \omega_{nk;j}^2 \, v_{nk;j}^{(1)}\, \, \hspace{0.5cm} \mbox{and} \hspace{0.5cm} (2) \, \, \, \, \, v_{nk;j}^{(N)}=0
\]
A good estimation of the discrete eigenvalues $\omega_{nk;j}^2$ is obtained through diagonalization of the $N\times N$ matrix in the left member of the linear system (\ref{erec1}). We show the eigenvalues of ${\cal H}^+$ for the lowest values of $n$ and $k$ in Table 1. In particular, we have implemented the algorithm (\ref{erec1}) in a Mathematica environment with a choice of two grids constructed respectively with $N=N_0=4000$ and $N=2N_0=8000$ grid points. The results achieved manifest a satisfactory reliability shown in the data displayed in Table 1. Different figures in the displayed results corresponding to the two previously mentioned meshes have been emphasized by enclosing them in parentheses. We remark that we have chosen generically the value $r_{\rm max}=15$ in the numerical scheme in such a way that the precision of the Laplacian operator in (\ref{erec1}) is of order ${\cal O}(\Delta x^2) \sim 3.4 \times 10^{-6}$ using the mesh with $2N_0=8000$ points. Special treatment is required for states whose eigenvalues are near the scattering threshold value $1$ where we choose $r_{\rm max}=50$ such that ${\cal O}(\Delta x^2) \sim 0.00004$. The reason is that eigenfunctions close to the scattering threshold exhibit quite slow decaying asymptotic behavior, which demands a greater value of $r_{\rm max}$.

\begin{table}[h]
\centering
\begin{tabular}{|c||c|c|c|c|c|} \hline
\multicolumn{6}{|c|}{\rule[-0.3cm]{0.0cm}{0.8cm}Eigenvalues of the discrete spectrum of ${\cal H}^+$} \\ \hline\hline
$n$ & $N$ & \rule[-0.3cm]{0.0cm}{0.8cm} $k=0$ & $k=1$ & $k=2$ & $k=3$ \\ \hline\hline
\multirow{2}{*}{1} & $N_0$ & \rule[-0.3cm]{0.0cm}{0.8cm} $(\omega_{10;1}^{\rm A})^2=0.777476(0)$ & - & - & \hspace{0.9cm} - \hspace{0.9cm} \\
 & $2N_0$ & \rule[-0.3cm]{0.0cm}{0.6cm} $(\omega_{10;1}^{\rm A})^2=0.777476(2)$ & - & - & \hspace{0.9cm} - \hspace{0.9cm} \\ \hline
\multirow{2}{*}{2}  & $N_0$ & \rule[-0.3cm]{0.0cm}{0.8cm}  $(\omega_{20;1}^{\rm A})^2=0.53859(69)$ & $(\omega_{21;1}^{\rm A})^2=0.97303(22)$ & - & - \\
 & $2N_0$ & \rule[-0.3cm]{0.0cm}{0.6cm}  $(\omega_{20;1}^{\rm A})^2=0.53859(71)$ & $(\omega_{21;1}^{\rm A})^2=0.97303(58)$ & - & - \\ \hline
\multirow{2}{*}{3} & $N_0$ & \rule[-0.3cm]{0.0cm}{0.8cm}  $(\omega_{30;1}^{\rm A})^2=0.402708(7)$ & $(\omega_{31;1}^{\rm A})^2=0.83025(60)$ & - & - \\
 & $2N_0$ & \rule[-0.3cm]{0.0cm}{0.6cm}  $(\omega_{30;1}^{\rm A})^2=0.402708(8)$ & $(\omega_{31;1}^{\rm A})^2=0.83025(73)$ & - & - \\ \hline

\multirow{4}{*}{4} & $N_0$ & \rule[-0.3cm]{0.0cm}{0.8cm} $(\omega_{40;1}^{\rm A})^2=0.319288(3)$ & $(\omega_{41;1}^{\rm A})^2=0.701767(2)$ & - & - \\
& $2 N_0$ & \rule[-0.3cm]{0.0cm}{0.6cm} $(\omega_{40;1}^{\rm A})^2=0.319288(4)$ & $(\omega_{41;1}^{\rm A})^2=0.701767(6)$ & - & - \\
 & $N_0$ & \rule[-0.3cm]{0.0cm}{0.8cm} $(\omega_{40;2}^{\rm A})^2=0.98835(24)$ &   &   &   \\
& $2 N_0$ & \rule[-0.3cm]{0.0cm}{0.6cm} $(\omega_{40;2}^{\rm A})^2=0.98835(36)$ &   &   &   \\ \hline

\multirow{4}{*}{5} & $N_0$ & \rule[-0.3cm]{0.0cm}{0.8cm} $(\omega_{50;1}^{\rm A})^2=0.263679(8)$ & $(\omega_{51;1}^{\rm A})^2=0.601272(6)$ & $(\omega_{52;1}^{\rm A})^2=0.94252(29)$ & - \\
& $2 N_0$ & \rule[-0.3cm]{0.0cm}{0.6cm} $(\omega_{50;1}^{\rm A})^2=0.263679(9)$ & $(\omega_{51;1}^{\rm A})^2=0.601272(9)$ & $(\omega_{52;1}^{\rm A})^2=0.94252(36)$ & - \\
 & $N_0$ & \rule[-0.3cm]{0.0cm}{0.8cm} $(\omega_{50;2}^{\rm A})^2=0.93845(63)$ &   &   &   \\
& $2 N_0$ & \rule[-0.3cm]{0.0cm}{0.6cm} $(\omega_{50;2}^{\rm A})^2=0.93845(87)$ &   &   &   \\ \hline

\end{tabular}
\caption{Numerical estimation of the discrete spectrum eigenvalues within the class A eigenfunctions $\xi_\lambda^{\rm A+}(\vec{x},n,k)$ of the second-order $n$-vortex small fluctuation operator ${\cal H}^+$. These results have been obtained using the algorithm (\ref{erec1}) with two different meshes of respectively $N_0=4000$ and $2N_0=8000$ points. Parentheses are used to point out the disagreeing figures arising from either the thicker
or the thinner meshes.}
\end{table}

There is a logical structure almost so rich as the structure of vortex zero modes. In general, we observe that the number of bound states increases with the magnetic flux $n$. In particular, for magnetic flux $n=1$, we find only one boson vector-vortex bound state. The eigenvalue of this state is $(\omega_{10,1}^A)^2=0.77747$, which arises for angular momentum $k=0$. For magnetic flux $n=2$, we find two boson vector-vortex bound states $(\omega_{20,1}^A)^2=0.53859$ and $(\omega_{21,1}^A)^2=0.97303$ whose Fourier wave numbers are $k=0$ and $k=1$ respectively. For magnetic flux $n=3$ the situation is completely similar to the case $n=2$ with eigenvalues $(\omega_{30,1}^A)^2=0.402708$ and $(\omega_{31,1}^A)^2=0.83025$. We find three bound states for the case $n=4$, two of them corresponding to the angular momentum $k=0$ with eigenvalues $(\omega_{40,1}^A)^2=0.319288$ and $(\omega_{40,2}^A)^2=0.98835$ while the remaining eigenvalue $(\omega_{41,1}^A)^2=0.701767$ corresponds to the value $k=1$. This last situation is also suitable for the case $n=5$ with the eigenvalues $(\omega_{50,1}^A)^2=0.263679$ and $(\omega_{50,2}^A)^2=0.938456$ and $(\omega_{51,1}^A)^2=0.601272$ although a fourth state with $k=2$ arises with eigenvalue $(\omega_{52,1}^A)^2=0.94252$.

Table 2 summarizes all the spectral information of the radial eigenvalue problem (\ref{ode55}) analyzed in this Section for vortex solutions with vorticity $n=5$. In this Table we display the effective potentials $V_{\rm eff}(r)=\frac{k^2}{r^2}+f_n^2(r)$ (blue lines), the boson-vortex bound states energies (red lines) and the radial eigenfunction profile for several values of $n$ (arrayed in different rows) and $k$ (arrayed in different columns). Two types of behavior are distinguished: (a) If $k=0$, we have a central potential with a minimum at the origin that tends to $1$ at $r\to \infty$. Above this threshold, $\omega_\lambda^2>1$, there is no question about the existence of scattering states in the spectrum of ${\cal H}^+$, but we also find one boson-$n$-vortex bound state for $n=1,2,3$ with an eigenvalue below the scattering threshold decreasing with $n$. For the $n=4$ and $n=5$ BPS vortices, a second boson-vortex bound state appear. (b) If $k>0$ the centrifugal barrier in the effective potential gives rise to a hard core but there is still a minimum of the potential away from the core if $k$ is small enough. All together we observe a shallower well
(that becomes a wall for a big enough $k$) that tends to infinity at $r=0$ and to $1$ when $r\to \infty$. In the second, third, and fourth columns of Table 1 it is observed that when $k$ grows we need a higher magnetic flux $n$ for finding bound states.

\begin{table}[ht]
\centering\begin{tabular}{|c|cccc|} \hline
\multicolumn{5}{|c|}{\rule[-0.3cm]{0.0cm}{0.8cm} Eigenvalues of the discrete spectrum of ${\cal H}^+$ displayed on the potential wells $V_{\rm eff}(\vec{x},n,k)$} \\ \hline\hline
$n$ & $k=0$ & $k=1$ &  $k=2$ &  $k=3$ \\ \hline\hline
\rotatebox{90}{\hspace{1cm}$n=1$} &\includegraphics[width=3.5cm]{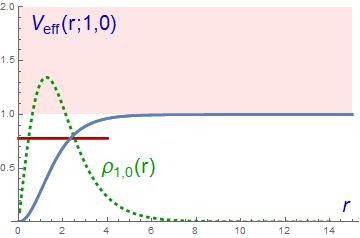} & \includegraphics[width=3.5cm]{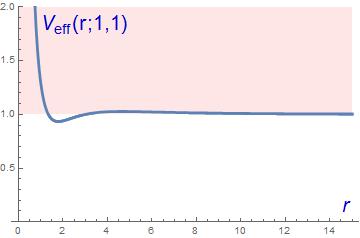} & & \\
\rotatebox{90}{\hspace{1cm}$n=2$} &\includegraphics[width=3.5cm]{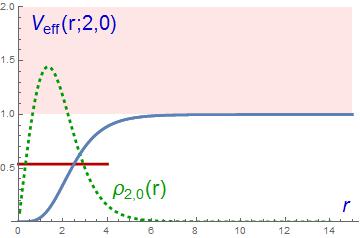} & \includegraphics[width=3.5cm]{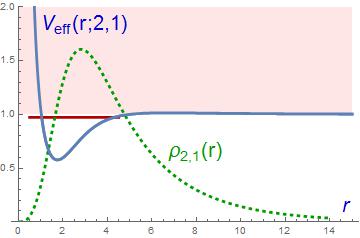} & \includegraphics[width=3.5cm]{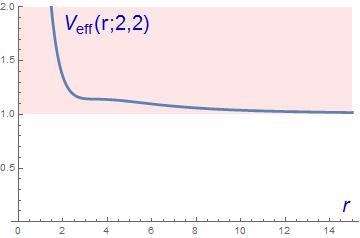} & \\
\rotatebox{90}{\hspace{1cm}$n=3$} &\includegraphics[width=3.5cm]{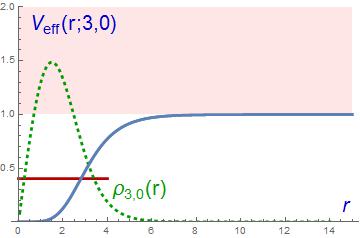} & \includegraphics[width=3.5cm]{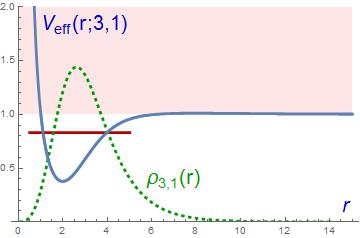} & \includegraphics[width=3.5cm]{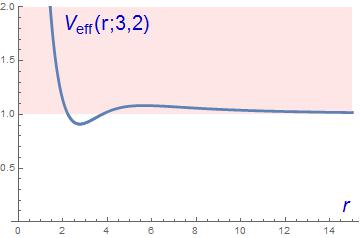} & \\
\rotatebox{90}{\hspace{1cm}$n=4$} &\includegraphics[width=3.5cm]{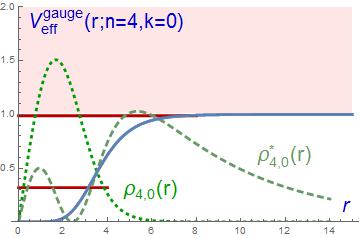} & \includegraphics[width=3.5cm]{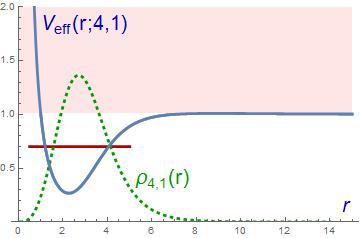} & \includegraphics[width=3.5cm]{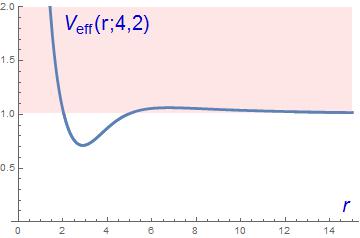} & \\
\rotatebox{90}{\hspace{1cm}$n=5$} &\includegraphics[width=3.5cm]{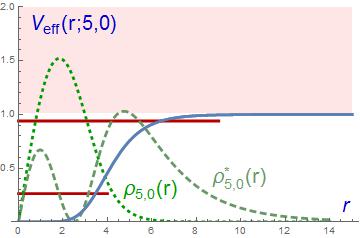} & \includegraphics[width=3.5cm]{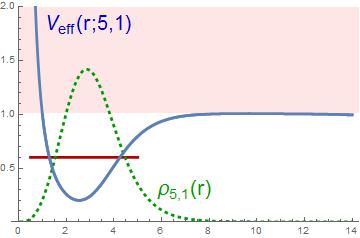} & \includegraphics[width=3.5cm]{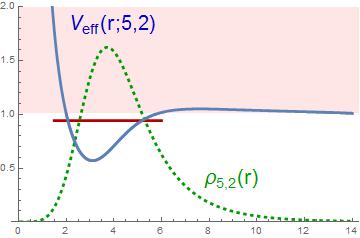} & \includegraphics[width=3.5cm]{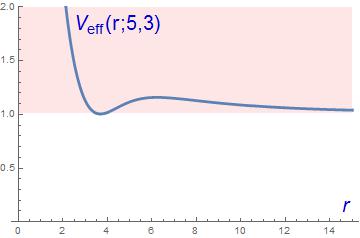} \\ \hline
\end{tabular}
\caption{Graphical representation of the effective potential wells/walls $V_{\rm eff}(\vec{x},n,k)$ in the differential Schr$\ddot{\rm o}$dinger equation (\ref{ode55}) for the lower values of $n$ and $k$. We superimpose the lower positive eigenvalues in the pure point ${\cal H}^+$-spectrum on these Figures.}
\end{table}

\subsubsection{BPS vortex bound state wave functions}

The boson-vortex bound state eigenfunctions are found by simulating the ODE system (\ref{ode55}) by the finite-difference equations system (\ref{erec1}) and searching for the eigenvectors. In Table 2 the radial form factors characterizing the positive eigenfunctions of ${\cal H}^\pm$ are plotted for each eigenvalue. The information obtained from these $v_{nk}(r)$ radial wave functions plugged into formulae (\ref{excitedmode1}) or (\ref{excitedmode2}) unveils the details of the bound state eigenfunctions of ${\cal H}^+$ of class A. Although these wave functions describe scalar and vector bosons trapped at the vortex core, we shall call them boson vector-vortex bound states because they come from the positive eigenfunctions of ${\cal H}^-$ containing only vector particles.

Properties of the boson vector-vortex bound states supporting $n=5$ quanta of magnetic flux are shown in Figure 2 in a Table format. Indeed we shall use this case as a paradigmatic example of the discrete bound state spectrum displayed in Table 1 because the qualitative behavior of the eigenfunctions carrying identical angular momentum $k$ is similar for every vorticity $n$.

In the first row of Figure 2 the scalar $\varphi^{A+}$ and vector $a^{A+}$ fluctuations corresponding to the lowest eigenvalue $(\omega_{50;1}^{\rm A})^2=0.263679$ are plotted, as well as the perturbed fields $\psi+\epsilon \, \varphi^{A+}$ and $V+\epsilon \, a^{A+}$ corresponding to the bound state positive fluctuation $\xi^{A+}_{1}(\vec{x};5,0)$ of a BPS $5$-vortex centered at the origin. It is interesting to examine the details: (1) Both $\varphi^{A+}(\vec{x};5,0)$ and $a^{A+}(\vec{x};5,0)$ have zeroes at the origin, grow in the middle region and tend to zero again at $r=\infty$. The BPS $5$-vortex does not vary under this $k=0$ fluctuation at its center, but grows at the core's middle, and remain fixed again near infinity, see Figure 2 (first row). (2) We see this instantaneous picture as an \lq\lq inflationary\rq\rq process, a grow of the vortex density at middle distances from the origin. But it is in fact an oscillatory process of inflation/deflation due to the temporal dependence induced by the periodic time-dependent term: $\cos \omega_{50,1} t$ \footnote{It is enlightening to compare this bound state with the excited bound state of a meson trapped by $\lambda\phi^4$-kink: $\phi_K(x)=\tanh x$ and
$\varphi_3(x)=\frac{\sinh x}{\cosh^2 x}$, $\omega_3^2=3$. Clearly, the kink remains unchanged at $x=0$, increases at $x >0$, but diminishes when $x<0$, and diminishes until $\phi \simeq 1$, far away from the kink center on the right, but increases until $\phi\simeq -1$ if $x\ll -1$. All this happens at $t=t_0$ but this frozen picture oscillates in time, with frequency $\omega_3=\sqrt{3}$.}.

\begin{figure}[H]
\begin{tabular}{c|cccc}
 & 1. $\varphi^{A+}(\vec{x})$ & 2. $\psi(\vec{x})+\epsilon \, \varphi^{A+}(\vec{x})$ & 3. $a^{A+}(\vec{x})$ &  4. $V(\vec{x}) + \epsilon \,a^{A+}(\vec{x})$ \\ \hline
\rotatebox{90}{\hspace{0.8cm} 9. $\xi_{1}^{A+}(\vec{x};5,0)$} &\includegraphics[width=3.5cm]{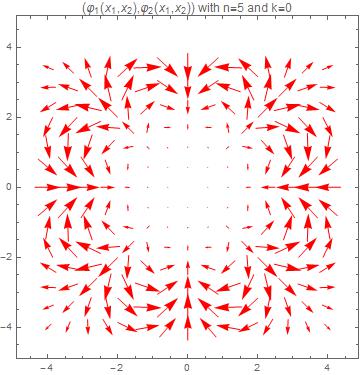} & \includegraphics[width=3.5cm]{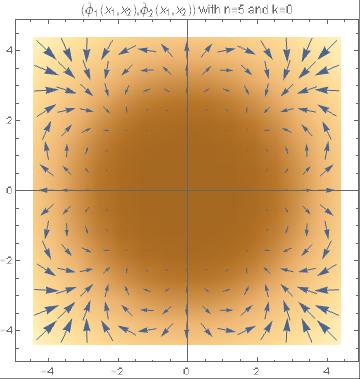} & \includegraphics[width=3.5cm]{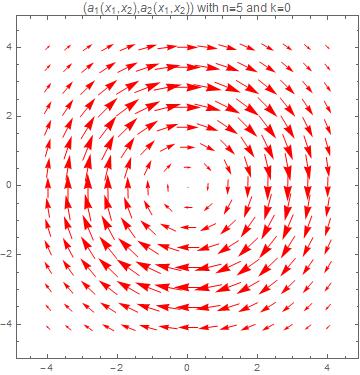} & \includegraphics[width=3.5cm]{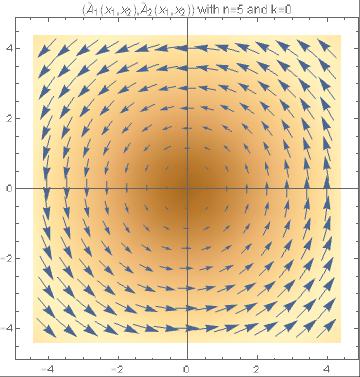} \\
\rotatebox{90}{\hspace{0.8cm} 10. $\xi_{2}^{A+}(\vec{x};5,0)$} &\includegraphics[width=3.5cm]{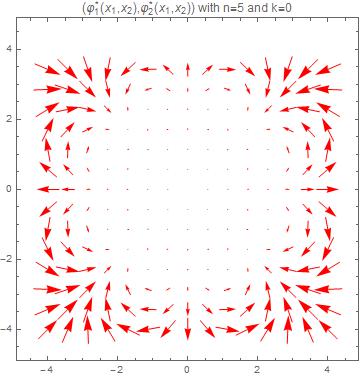} & \includegraphics[width=3.5cm]{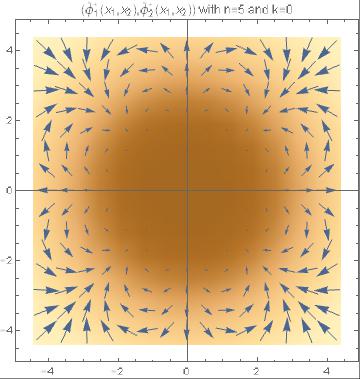} & \includegraphics[width=3.5cm]{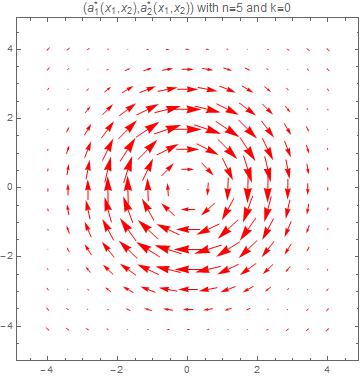} & \includegraphics[width=3.5cm]{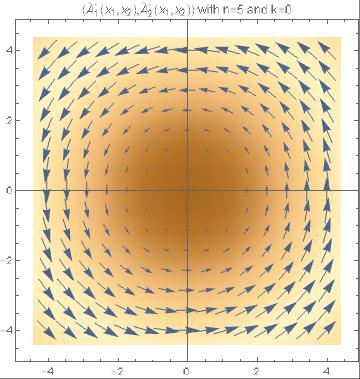} \\
\rotatebox{90}{\hspace{0.8cm} 11. $\xi_{1}^{A+}(\vec{x};5,1)$} &\includegraphics[width=3.5cm]{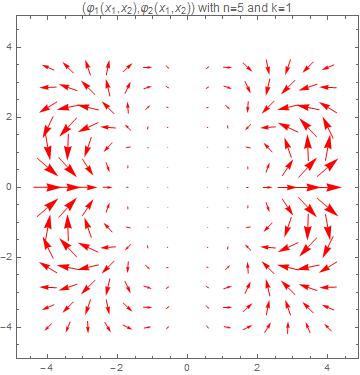} & \includegraphics[width=3.5cm]{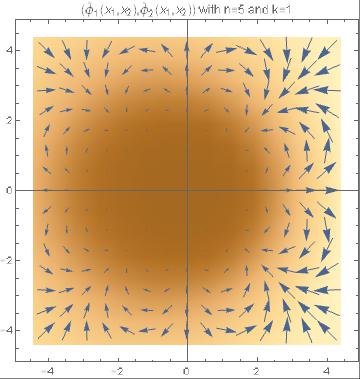} & \includegraphics[width=3.5cm]{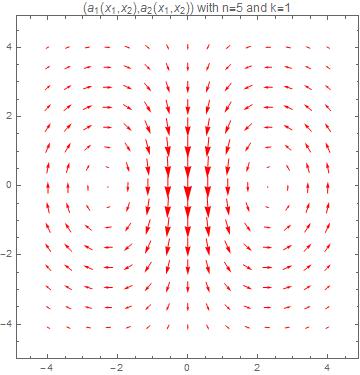} & \includegraphics[width=3.5cm]{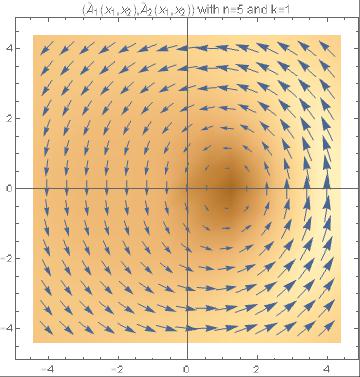} \\
\rotatebox{90}{\hspace{0.8cm} 12. $\xi_{1}^{A+}(\vec{x};5,2)$} &\includegraphics[width=3.5cm]{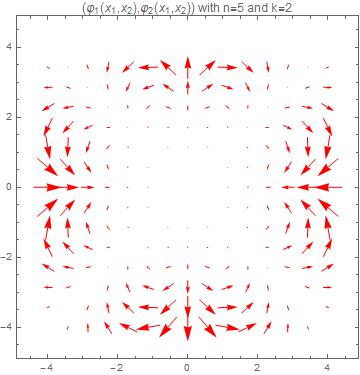} & \includegraphics[width=3.5cm]{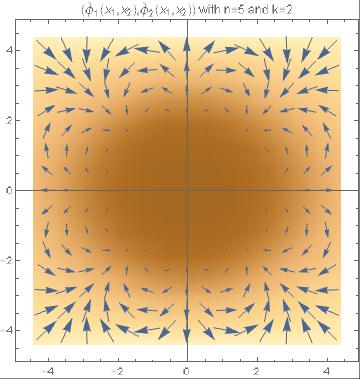} & \includegraphics[width=3.5cm]{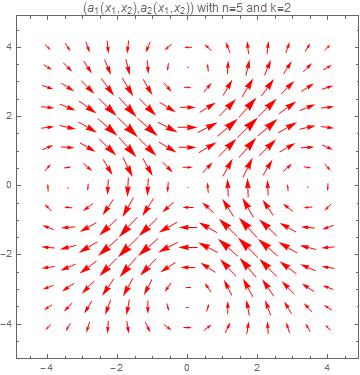} & \includegraphics[width=3.5cm]{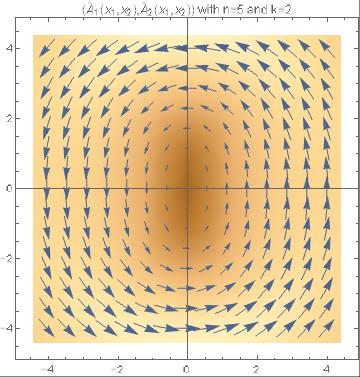} \\
\end{tabular}
\caption{Vector boson-vortex bound states $\xi_1^{A+}(\vec{x};5,0)$, $\xi_2^{A+}(\vec{x};5,0)$, $\xi_1^{A+}(\vec{x};5,1)$ and
$\xi_1^{A+}(\vec{x};5,2)$ on a BPS $5$-vortex and the self-dual $5$-vortex fields perturbed by this fluctuation.}
\end{figure}

The next excited vector-boson bound state in the discrete ${\cal H}^+$-spectrum is the eigenmode with Fourier wave number $k=1$ whose eigenvalue is $(\omega_{51,1}^A)^2=0.601272$. The radial form factor $v_{51,1}(r)$ corresponding to this eigenfunction is shown in Table 2 (last row) while the $\xi^{A+}_{1}(\vec{x};5,1)$ vector-vortex bound state is graphically depicted in the third row of Figure 2 (third row). We observe a new class of fluctuations. The scalar fluctuation $\varphi^{A+}(\vec{x};5,1)$ is stronger in a strip enclosing the $x_1$-axis when it differs appreciably from zero for intermediate distances from the origin. The net effect on $\psi(\vec{x},2)$ is the squeezing of the $2$-vortex along the $x_1$-axis both from the left and from the right, but less intensely along the $x_2$-axis in both senses, see Figure 2 (third row). Capturing the scalar and vector bosons in this bound state the BPS rotationally symmetric $5$-vortex loses symmetry and gets thinner along the $x_1$-axis. Of course, there is an oscillation of this static picture, with frequency $\omega_{51,1}^A=\sqrt{0.601272}$ with a certain dipolar character. A similar process emerges for the vector self-dual $5$-vortex field $V(\vec{x},5)$ and the corresponding fluctuation $a^{A+}(\vec{x},5,1)$, see Figure 2.

In addition to the first mentioned fluctuation mode there exists other bound state in the ${\cal H}^+$ spectrum with angular momentum $k=0$ whose eigenvalue is $(\omega_{50,2}^{\rm A})^2=0.93845$. The perturbed fields behave following a similar pattern than that of the previously described $k=0$-eigenmode, see Figure 2 (second row), although the radial profile function involves a new node, see the probability function $\rho_{5,0}^*(r)$ in the last row of Table 2.

The highest eigenvalue in the discrete ${\cal H}^+$-spectrum takes the value $(\omega_{52,1}^A)^2=0.94252$. This peculiar state $\xi_{1}^{\rm A+}(\vec{x};5,2)$ is described in the last row of Figure 2. Focusing our attention in $\varphi^{\rm A+}(\vec{x};5,2)$ we observe that the perturbation pressures the scalar field density concentrated in a disk inwards along the $x_1$-axis but outwards along the $x_2$-axis, inducing a quadrupolar density distribution oscillating in time.

Animations illustrating the previously mentioned behavior of the fluctuation eigenmodes together with the energy density and magnetic field associated with them can be downloaded at the web page http://campus.usal.es/~mpg/General/Mathematicatools.

\section{Brief summary and outlook}

An ortho-normal basis of BPS cylindrically symmetric $n$-vortex zero modes in the kernel of the matrix second-order PED operator
${\cal H}^+$ of fluctuation has been constructed and their mathematical and physical properties thoroughly described. Several positive
normalizable eigenfunctions, bound states, of this Hessian operator, as well as their corresponding eigenvalues, have been identified
within an unexpected level of precision. This novel type of BPS cylindrically symmetric positive vector boson-vortex bound state fluctuations exhibits very intriguing properties and deserves further investigation.

The techniques and procedures used in this paper suggest that the concepts developed and the results attained in this basic superconducting setting will work in more sophisticated superconducting media provided that a BPS or self-dual structure is available. The first system of this type that comes to mind is the Chern-Simons-Higgs model, see \cite{Jackiw1}. One expects a similar structure of the self-dual Chern-Simons vortex zero mode fluctuation, demanding only an appropriate radial form factor $h_{nk}(r)$ to be estimated by the same method.
More dubious, however, is the prospect of finding some boson-vortex bound state in this necessarily planar model. Another possible scenario to play around with these ideas is the broad domain of $U(1)$-gauged massive non-linear sigma models, see \cite{Nitta,Wifredo}. Severely constrained by the possibility of building extended ${\cal N}=2$ supersymmetry on these systems, the main difficulty in investigating vortex zero modes and bound states in this framework appears to be the precise development of an index theorem in this context. Even more tantalizing, the development of a similar analysis about the zero modes of the BPS defects in the system discussed in \cite{Nitta1} seems to be plausible, even though the high derivative terms look harmful.

\end{document}